# A direct Jacobian total Lagrangian explicit dynamics finite element algorithm for real-time simulation of hyperelastic materials

Jinao Zhang

Department of Mechanical and Aerospace Engineering, Monash University, Clayton, Victoria, Australia

**Abstract**

This paper presents a novel direct Jacobian total Lagrangian explicit dynamics (DJ-TLED) finite element algorithm for real-time nonlinear mechanics simulation. The nodal force contributions are expressed using only the Jacobian operator, instead of the deformation gradient tensor and finite deformation tensor, for fewer computational operations at run-time. Owing to this proposed Jacobian formulation, novel expressions are developed for strain invariants and constant components, which are also based on the Jacobian operator. Results show that the proposed DJ-TLED consumed between 0.70x and 0.88x CPU solution times compared to state-of-the-art TLED and achieved up to 121.72x and 94.26x speed improvements in tetrahedral and hexahedral meshes, respectively, using GPU acceleration. Compared to TLED, the most notable difference is that the notions of stress and strain are not explicitly visible in the proposed DJ-TLED but embedded implicitly in the formulation of nodal forces. Such a force formulation can be beneficial for fast deformation computation and can be particularly useful if the displacement field is of primary interest, which is demonstrated using a neurosurgical simulation of brain deformations for image-guided neurosurgery. The present work contributes towards a comprehensive DJ-TLED algorithm concerning isotropic and anisotropic hyperelastic constitutive models and GPU implementation. The source code is available at https://github.com/jinaojakezhang/DJTLED.

Corresponding author at: Department of Mechanical and Aerospace Engineering, Monash University, Wellington Road, Clayton, VIC 3800, Australia.
*E-mail addresses*: jinao.zhang@monash.edu; jinao.zhang@hotmail.com

## 1. Introduction

Real-time and accurate simulation of nonlinear deformations of mechanical problems is becoming increasingly important in many interactive engineering applications. It can be used to allow interactive design, analysis and optimisation in computer-aided design [1] and manufacturing [2], to enable real-time control of soft robots [3], to permit real-time pose estimation for soft sensors [4], and to achieve user interactions with deformable bodies in virtual-reality and augmented-reality environments [5]. The real-time and accurate simulation of nonlinear deformations is also essential to many biomedical applications such as patient-specific whole-body image registration [6], intraoperative brain-shift compensation [7], virtual training for electrocardiology procedures [8], and real-time surgical simulation of cataract surgery, laparoscopic surgery and tumour removal [9].

Currently, many of the reported numerical algorithms are mainly focused on the improvement of numerical accuracy [10,11] and convergence [12], with fewer considerations on computational efficiency. These algorithms utilise sophisticated constitutive formulations and computing procedures to achieve a very high order of accuracy, but the solutions are often computationally expensive to obtain. In contrast, interactive applications require the computation to be done in a short time to allow for immediate visualisation of the results, using consumer-level computing hardware. These relatively conflicting requirements (real-time and accurate) lead to a challenging task in which the performance improvement on one aspect (such as real-time computation) is often obtained to the detriment of the other (such as numerical accuracy); hence, a balance between the two requirements is often made.

This work is focused on the improvement of computational efficiency for solutions of finite-strain problems, for which nonlinear continuum mechanics is the fundamental basis. The most significant challenge arises from the solution of the nonlinear governing equation where geometric and material nonlinearities are involved. The nonlinear system of discretised equations is typically solved based on the Newton-Raphson's procedure through a sequence of solutions of a linearised system of equations which can be solved by either (i) direct inverse or factorisation of the system matrix or (ii) iteratively solving a system of algebraic equations based on an initial estimate. However, both procedures are computationally expensive.

Various methods were studied for real-time geometric and material nonlinear deformation computation. Some representative mesh-based methods include total Lagrangian explicit dynamics (TLED) finite element algorithm [13], hyperelastic mass links (HEML) [14], and multiplicative Jacobian energy decomposition [15]. Some representative meshless methods include meshless TLED [16], point collocation method of finite spheres [17], and smooth particle hydrodynamics [18]. Some recently reported methodologies were focused on model order reduction [19,20] and deep learning [21] to embed high dimensional problems in a smaller subspace. Despite a significant reduction in model dimensions, these methods typically require the training of the reduced models or the neural networks, and the computed deformations are sensitive to the quality of data used for training.

The current state-of-the-art real-time nonlinear finite element algorithm is often based on the TLED formulation, which employs the total Lagrangian description, explicit dynamics, and low-order finite elements for a balance of numerical accuracy with very efficient computation.

The displacement-based finite element analysis uses updated Lagrangian (UL) or total Lagrangian (TL) to describe the frame of reference [22]. UL applies an incremental strain description for variable values which requires re-calculation of spatial derivatives at each time step. In contrast, TL considers variables referred to the reference

(undeformed) configuration so that the spatial derivatives only need to be computed once, and the strains lead to correct results after a load cycle without error accumulation [23]. It was reported that TL took 2.1 $ms$, compared to 10.6 $ms$ by UL, for a time step in the computation of an ellipsoid indentation [24].

Explicit dynamics use the lumped mass formulation and explicit time integration for very efficient computation and is well-suited for parallel computation. The internal and external loads and mass are lumped to the nodes, leading to block diagonalised mass and damping matrices that allow for element-level computation [13], without the need for the assemblage of the global system matrices. The explicit time integration allows for direct computation of state variables in future time points based on their values at the current time point only, without the need for solving a system of equations in the implicit integration. It was reported that the implicit scheme consumed at least one order of magnitude more time than the explicit counterpart [25].

Low-order finite elements are used in TLED. It was discussed that the finite element mesh needs to use low-order finite elements that are not computationally intensive to meet the computation time (real-time) requirement [26]. This means that the 4-node tetrahedral and the 8-node hexahedral (brick) finite elements using the polynomials of degree one $P_1$ (linear) for interpolation functions (also called shape functions) and 1-point Gauss integration are used.

This paper is focused on further improving the computational efficiency of state-of-the-art TLED by developing a novel direct Jacobian TLED (DJ-TLED) algorithm. The motivation is to enable shorter computation time and larger model sizes to be simulated in nonlinear mechanics problems. The proposed DJ-TLED is based on a reformulation of TLED using only the Jacobian operator, and hence it is referred to as direct Jacobian TLED. The benefit of this Jacobian-only formulation is that it leads to extra time-invariant components to be precomputed and updates only the Jacobian operator at run-time, instead of updating the full deformation tensor in TLED, resulting in fewer online computational operations. The novelties of the proposed method are to express all formulations using solely the Jacobian operator, including the formulations for nodal force contributions, strain invariants and constant tensors for efficient computation. The following contributions are presented:

(i) proposed DJ-TLED formulation (Section 2.2),
(ii) Jacobian formulation of strain invariants for isotropic hyperelastic materials (Section 2.3.1),
(iii) Jacobian formulation of pseudo-invariants for anisotropic hyperelastic materials (Section 2.3.2),
(iv) formulation of new time-invariant tensors for isotropic hyperelastic materials (Section 2.4.1),
(v) formulation of new time-invariant tensors for anisotropic hyperelastic materials (Section 2.4.2), and
(vi) Graphics Processing Unit (GPU) implementation (Section 4).

Results show that the proposed DJ-TLED consumed between 0.70x and 0.88x CPU solution times compared to TLED under the same conditions using the test computing hardware. Using GPU acceleration, it achieved speed improvements of up to 121.72x and 94.26x in the 4-node tetrahedral mesh and the 8-node hexahedral mesh, respectively, over CPU execution. The translational benefits of the proposed DJ-TLED are demonstrated using a neurosurgical simulation of brain deformations (Section 6).

The remainder of this paper is organised as follows: the constitutive framework of the proposed DJ-TLED is presented in Section 2; examples of nodal force formulations are given in Section 3; GPU implementation is presented in Section 4; Section 5 presents algorithm verification and performance evaluation; Section 6 demonstrates an application of DJ-TLED to neurosurgical simulation; discussions are presented in Section 7, and finally, the paper concludes in Section 8 with future improvements.

## 2. Methods

The constitutive framework of the proposed DJ-TLED finite element algorithm is presented. Formulations are developed for both isotropic and anisotropic hyperelastic materials.

*2.1 Preliminaries*

Consider a continuum body $\mathcal{B}$ composed of a set of material points $P \in \mathcal{B}$ that moves in the three-dimensional Euclidean space from one instant of time to another, it occupies a continuous sequence of geometrical regions denoted by $^0\Omega, ..., \ ^t\Omega$ where $^0\Omega$ is referred to as the reference (undeformed) configuration and $^t\Omega$ the current (deformed) configuration of the body $\mathcal{B}$. We consider that the body $\mathcal{B}$ can change its shape which is said to be deformable. The left superscript denotes the instant of time $t$ of the configuration in which a quantity occurs.

A point $x \in P$ is identified by the position vectors $^0\mathbf{x} \in \ ^0\Omega$ and $^t\mathbf{x} \in \ ^t\Omega$, and they are related by the kinematics $^t\mathbf{u}: \ ^0\mathbf{x} \to \ ^t\mathbf{x}: \ ^0\Omega \times \mathbb{R}_0^+ \to \ ^t\Omega$ described by the motion of the body $\mathcal{B}$. Using finite element method, the body $\mathcal{B}$ is divided into finite elements forming a finite element mesh that conforms to the continuum in a discrete manner.

### 2.1.1 Nodal force computation in TLED finite element algorithm

TLED finite element algorithm [13] employs the TL description for the frame of reference and explicit dynamics for the nodal mass and time integration. Due to using a lumped (diagonal) mass matrix, the need for an assemblage of the stiffness matrix of the entire model is eliminated, and the computation is performed at the element level.

For a given element at time $t$, the nodal force contributions $^t_0\mathbf{F}$ due to stresses are computed by

$$^t_0\mathbf{F} = \int_{^0V} ^t_0\mathbf{B}_L^T \ ^t_0\hat{\mathbf{S}} \, d \ ^0V \tag{1}$$

where $^t_0\mathbf{B}_L$ is the full strain-displacement transformation matrix, $^t_0\hat{\mathbf{S}}$ is the vector of second Piola-Kirchhoff stresses at integration points (recognising the tensor's symmetry), and $^0V$ is the initial volume of the element. The left subscript denotes the configuration with respect to which the quantity is measured. $(\cdot)^T$ denotes the transpose of a matrix. Notation of Bathe [22] is used. A detailed description of TLED finite element algorithm is presented in Ref. [13].

### 2.1.2 Jacobian operator

The Jacobian operator relates the natural coordinate derivatives to the local coordinate derivatives [22]. In the isoparametric finite element formulation, the element coordinates $\mathbf{x}(x, y, z)$ and displacements $\mathbf{u}(u_x, u_y, u_z)$ are expressed in the form of interpolation functions defined in the natural coordinate system $\boldsymbol{\xi}(\xi, \eta, \zeta)$ of the element where each variable vary from -1 to +1. To evaluate the required derivatives, the following expression is used:

$$\frac{\partial}{\partial \boldsymbol{\xi}} = \mathbf{J} \frac{\partial}{\partial \mathbf{x}} \qquad (2)$$

where **J** is the Jacobian operator expressed by

$$\mathbf{J} = \begin{bmatrix} \frac{\partial x}{\partial \xi} & \frac{\partial y}{\partial \xi} & \frac{\partial z}{\partial \xi} \\ \frac{\partial x}{\partial \eta} & \frac{\partial y}{\partial \eta} & \frac{\partial z}{\partial \eta} \\ \frac{\partial x}{\partial \zeta} & \frac{\partial y}{\partial \zeta} & \frac{\partial z}{\partial \zeta} \end{bmatrix} \qquad (3)$$

The inverse of **J** exists provided that there is a unique correspondence between the natural and the local coordinates of the element.

### 2.2 Proposed direct Jacobian TLED (DJ-TLED) finite element algorithm

The proposed DJ-TLED is motivated by the need for achieving higher computational efficiency than state-of-the-art TLED while maintaining the same order of numerical accuracy for real-time (interactive) applications, concerning finite-strain deformable modelling where both geometric and material nonlinearities are involved.

To this end, the nodal force contributions ${}_0^t\mathbf{F}$ in TLED are formulated in terms of the Jacobian operator **J** entirely, in order to reduce computational complexity. Fig. 1 illustrates the computation in the proposed DJ-TLED. The novelties arise from using solely the Jacobian operator to express hyperelastic nodal forces and finite-strain components for efficient online computation.

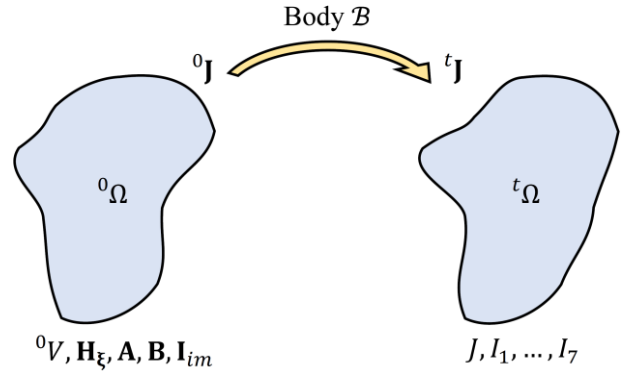

**Fig. 1.** Reference ${}^0\Omega$ and current ${}^t\Omega$ configurations of the body $\mathcal{B}$: the proposed DJ-TLED is based on the Jacobian operator **J** for deformation computation where the initial element volume ${}^0V$, interpolation function natural derivatives $\mathbf{H}_\xi$, structural matrices **A**, **B** indicating anisotropy, and new time-invariant tensors $\mathbf{I}_{im}$ ($i = 1, 2, 4, 5, 6, 7$) are defined in ${}^0\Omega$, and volume ratio $J$ and strain and pseudo-invariants $I_1, \ldots, I_7$ are defined in ${}^t\Omega$.

Consider the matrix form of second Piola-Kirchhoff stresses ${}_0^t\mathbf{S}$, the nodal force contributions ${}_0^t\mathbf{F}$ can be written as [26]

$$ {}_0^t\mathbf{F} = {}_0^t\mathbf{X} {}_0^t\mathbf{S}\ {}^0\mathbf{B}\ {}^0V \qquad (4) $$

where ${}_0^t\mathbf{X}$ denotes the deformation gradient which is a two-point tensor relating points in two distinct configurations (${}^0\Omega$ and ${}^t\Omega$ in this case), and ${}^0\mathbf{B}$ is the strain-displacement transformation matrix.

By using the Jacobian operator $\mathbf{J}$ (Eq. (3)) at time 0 and $t$, the deformation gradient tensor ${}_0^t\mathbf{X}$ can be expressed as

$$ {}_0^t\mathbf{X} = \begin{bmatrix} \dfrac{\partial\ {}^tx}{\partial\ {}^0x} & \dfrac{\partial\ {}^tx}{\partial\ {}^0y} & \dfrac{\partial\ {}^tx}{\partial\ {}^0z} \\ \dfrac{\partial\ {}^ty}{\partial\ {}^0x} & \dfrac{\partial\ {}^ty}{\partial\ {}^0y} & \dfrac{\partial\ {}^ty}{\partial\ {}^0z} \\ \dfrac{\partial\ {}^tz}{\partial\ {}^0x} & \dfrac{\partial\ {}^tz}{\partial\ {}^0y} & \dfrac{\partial\ {}^tz}{\partial\ {}^0z} \end{bmatrix} = \dfrac{\partial\ {}^t\mathbf{x}}{\partial\boldsymbol{\xi}}\dfrac{\partial\boldsymbol{\xi}}{\partial\ {}^0\mathbf{x}} = {}^t\mathbf{J}^T\ {}^0\mathbf{J}^{-T} \qquad (5) $$

In addition to the physical meaning of ${}_0^t\mathbf{X}$ which indicates the differentiations of the local coordinates ${}^t\mathbf{x}$ with respect to ${}^0\mathbf{x}$, the above ${}_0^t\mathbf{X} = {}^t\mathbf{J}^T\ {}^0\mathbf{J}^{-T}$ is expressed in the form of multiplicative decomposition in terms of the Jacobian operator at time 0 and $t$, based on the common intermediate natural coordinates $\boldsymbol{\xi}$.

The strain-displacement transformation matrix ${}^0\mathbf{B}$ can be expressed in terms of the Jacobian operator at time 0 as

$$ {}^0\mathbf{B} = {}^0\mathbf{J}^{-1}\mathbf{H}_{\boldsymbol{\xi}} \qquad (6) $$

where $\mathbf{H}_{\boldsymbol{\xi}} = \dfrac{\partial\mathbf{H}}{\partial\boldsymbol{\xi}}$ is defined as the differentiation of the matrix of interpolation functions $\mathbf{H}$ with respect to the natural coordinates $\boldsymbol{\xi}$ of the element, which can be precomputed; it is assumed that the same interpolation functions are used in finite elements in ${}^0\Omega$ and ${}^t\Omega$, and their natural derivatives $\mathbf{H}_{\boldsymbol{\xi}}$ are constant throughout the simulation. At any time $t$, the full, complete strain-displacement matrix ${}_0^t\mathbf{B}_L$ is obtained by transforming the stationary matrix ${}^0\mathbf{B}$ using the deformation gradient ${}_0^t\mathbf{X}$ (considered in Eq. (4)), which accounts for the initial displacement effect [13].

The next step is to express the second Piola-Kirchhoff stress tensor ${}_0^t\mathbf{S}$ by the Jacobian operator $\mathbf{J}$. Consider the strain energy density function $\Psi$ expressed in an uncoupled form $\Psi({}_0^t\mathbf{C}) = \Psi({}_0^t\bar{\mathbf{C}}) + \Psi(J)$ consisted of an isochoric part $\Psi({}_0^t\bar{\mathbf{C}})$ and a volumetric part $\Psi(J)$, the second Piola-Kirchhoff stress tensor ${}_0^t\mathbf{S}$ is expressed by

$$ {}_0^t\mathbf{S} = 2\left(\mathrm{DEV}\left[\dfrac{\partial\Psi({}_0^t\bar{\mathbf{C}})}{\partial\ {}_0^t\mathbf{C}}\right] + \dfrac{\partial\Psi(J)}{\partial\ {}_0^t\mathbf{C}}\right) \qquad (7) $$

where ${}_0^t\mathbf{C}$ is the right Cauchy-Green (deformation) tensor given by ${}_0^t\mathbf{C} = {}_0^t\mathbf{X}^T\ {}_0^t\mathbf{X}$, ${}_0^t\bar{\mathbf{C}} = J^{-2/3}\ {}_0^t\mathbf{C}$, $J$ is the determinant of the deformation gradient given by $J = \det({}_0^t\mathbf{X})$ where $\det(\cdot)$ denotes the determinant of a matrix, and $\mathrm{DEV}[\cdot]$ is the deviatoric projection operator given by $\mathrm{DEV}[\cdot] \equiv [\cdot] - \dfrac{1}{3}([\cdot]:{}_0^t\bar{\mathbf{C}}){}_0^t\bar{\mathbf{C}}^{-1}$.

For the purpose of simplicity, consider an isotropic hyperelastic material model where the isochoric part $\Psi({}_0^t\bar{\mathbf{C}})$ depends on the principal invariant $\bar{I}_1$ of ${}_0^t\bar{\mathbf{C}}$, i.e., $\Psi({}_0^t\mathbf{C}) = \Psi(\bar{I}_1) + \Psi(J)$. Without loss of generality, the dependences on other strain invariants and pseudo-invariants due to anisotropy are presented in the next section

(Section 2.3). The second Piola-Kirchhoff stress tensor ${}_0^t\mathbf{S}$ corresponding to $\Psi({}_0^t\mathbf{C}) = \Psi(\bar{I}_1) + \Psi(J)$ is expressed as

$$ {}_0^t\mathbf{S} = 2J^{-2/3}\frac{\partial \Psi}{\partial \bar{I}_1}\mathbf{I} + \left(-\frac{2}{3}\frac{\partial \Psi}{\partial \bar{I}_1}\bar{I}_1 + J\frac{\partial \Psi}{\partial J}\right){}_0^t\mathbf{C}^{-1} \tag{8}$$

where $\mathbf{I}$ is the identity matrix of the second rank.

By substituting the expressions of ${}_0^t\mathbf{X}$, ${}^0\mathbf{B}$ and ${}_0^t\mathbf{S}$ (Eqs. (5), (6) and (8)) in Eq. (4), the nodal force contributions ${}_0^t\mathbf{F}$ can be written as (see Appendix A)

$$ {}_0^t\mathbf{F} = {}_0^t\mathbf{X}{}_0^t\mathbf{S}\,{}^0\mathbf{B}\,{}^0V = \left(J^{-2/3}\,{}^t\mathbf{J}^T\frac{\partial \Psi}{\partial \bar{I}_1}\left(2\,{}^0V\,{}^0\mathbf{J}^{-T}\,{}^0\mathbf{J}^{-1}\right) + \left(-\frac{2}{3}\frac{\partial \Psi}{\partial \bar{I}_1}\bar{I}_1 + J\frac{\partial \Psi}{\partial J}\right){}^0V\,{}^t\mathbf{J}^{-1}\right)\mathbf{H}_\xi \tag{9}$$

In the proposed DJ-TLED, we denote the above equation by

$$ {}_0^t\mathbf{F} = \left(J^{-2/3}\,{}^t\mathbf{J}^T\frac{\partial \Psi}{\partial \bar{I}_1}\mathbf{I}_{1m} + \left(-\frac{2}{3}\frac{\partial \Psi}{\partial \bar{I}_1}\bar{I}_1 + J\frac{\partial \Psi}{\partial J}\right){}^0V\,{}^t\mathbf{J}^{-1}\right)\mathbf{H}_\xi \tag{10}$$

where $\mathbf{I}_{1m}$ is a constant time-invariant matrix defined by $\mathbf{I}_{1m} = 2\,{}^0V\,{}^0\mathbf{J}^{-T}\,{}^0\mathbf{J}^{-1}$ which can be precomputed. The other constant time-invariant components such as $\mathbf{I}_{2m}$, $\mathbf{I}_{4m}$, $\mathbf{I}_{5m}$, $\mathbf{I}_{6m}$ and $\mathbf{I}_{7m}$ for isotropic and anisotropic hyperelastic constitutive models are presented in Section 2.4, and they can also be precomputed.

The principal invariant $\bar{I}_1$ in the above equation can be computed from the proposed Jacobian formulation based on the Jacobian operator. The Jacobian formulations for strain invariants and pseudo-invariants are presented in Section 2.3.

As such, the nodal force contributions ${}_0^t\mathbf{F}$ in TLED are expressed in terms of the Jacobian operator $\mathbf{J}$ entirely, and hence we call the purposed methodology the direct Jacobian TLED (DJ-TLED). Precomputation can be performed for $\mathbf{I}_{1m}$, ${}^0V$ and $\mathbf{H}_\xi$; whereas $J$ and $\bar{I}_1$ are obtained from the Jacobian operator. Therefore, it needs to update only the Jacobian operator ${}^t\mathbf{J}$ at run-time, leading to a computationally more efficient algorithm than TLED. A comparison between DJ-TLED and TLED is presented in Section 7.

In the proposed DJ-TLED, the Jacobian operator $\mathbf{J}$ at any time $t$ can be conveniently computed by

$$ {}^t\mathbf{J} = {}^0\mathbf{J} + {}^t\mathbf{u}\mathbf{H}_\xi^T \tag{11}$$

and $J$ can be computed by

$$ J = \det({}_0^t\mathbf{X}) = \det({}^t\mathbf{J})\det({}^0\mathbf{J})^{-1} \tag{12}$$

where $\det({}^0\mathbf{J})$ can be precomputed.

At each time step, the global system of discretised equations to be solved is given by

$$ \mathbf{M}\ddot{\mathbf{U}} + \mathbf{D}\dot{\mathbf{U}} + \sum_{e=1}^{n}{}_0^t\mathbf{F}^{(e)} = \mathbf{R} \tag{13}$$

where $\mathbf{M}$ is the mass matrix (a lumped, diagonalised mass matrix is used), $\mathbf{D}$ is the damping matrix (a lumped, diagonalised mass-proportional damping, a special case of Rayleigh damping, is used), $\mathbf{R}$ is the vector of externally applied nodal forces, the dot notations $\dot{\mathbf{U}}$ and $\ddot{\mathbf{U}}$ denote the first- and second-order time derivatives of the global nodal displacements **U**, respectively, and ${}_0^t\mathbf{F}^{(e)}$ are the global nodal force contributions (such as Eq. (10)) due to stresses in element $e$.

The time integration in the temporal domain is performed using the central-difference scheme (a detailed description is presented in Ref. [27]). Due to explicit time integration, the time increment size $\Delta t$ must meet the Courant-Friedrichs-Lewy condition [28] for numerical stability, and the critical time step size is computed by

$$\Delta t \leq \frac{L_e}{c} \tag{14}$$

where $L_e$ is the smallest characteristic length of an element in the mesh, and $c$ is the dilatational wave speed.

*2.3 Proposed new formulations for strain invariants using Jacobian operator*

The independent strain invariants of the right Cauchy-Green tensor ${}_0^t\mathbf{C}$ are often used to express the strain energy density function $\Psi$ for the strain energy stored in the deformable continuum. In TLED, the strain invariants are obtained from ${}_0^t\mathbf{C}$ based on ${}_0^t\mathbf{C} = {}_0^t\mathbf{X}^T {}_0^t\mathbf{X}$. However, the proposed DJ-TLED does not require explicit computation of ${}_0^t\mathbf{X}$ and ${}_0^t\mathbf{C}$; instead, it uses only the Jacobian operator **J**. Hence, new formulations are developed to express the strain invariants using the Jacobian operator.

By using the Jacobian operator **J** at time $0$ and $t$, the right Cauchy-Green tensor ${}_0^t\mathbf{C}$ can be expressed as

$${}_0^t\mathbf{C} = {}_0^t\mathbf{X}^T {}_0^t\mathbf{X} = {}^0\mathbf{J}^{-1}\ {}^t\mathbf{J}\ {}^t\mathbf{J}^T\ {}^0\mathbf{J}^{-T} \tag{15}$$

which can be written in terms of individual components as

$$\begin{aligned}{}_0^t\mathbf{C} &= {}^tg_{11}\ {}^0\mathbf{J}^{-1}\begin{bmatrix}1&0&0\\0&0&0\\0&0&0\end{bmatrix}{}^0\mathbf{J}^{-T} + {}^tg_{22}\ {}^0\mathbf{J}^{-1}\begin{bmatrix}0&0&0\\0&1&0\\0&0&0\end{bmatrix}{}^0\mathbf{J}^{-T} + {}^tg_{33}\ {}^0\mathbf{J}^{-1}\begin{bmatrix}0&0&0\\0&0&0\\0&0&1\end{bmatrix}{}^0\mathbf{J}^{-T} \\ &+ {}^tg_{12}\ {}^0\mathbf{J}^{-1}\begin{bmatrix}0&1&0\\1&0&0\\0&0&0\end{bmatrix}{}^0\mathbf{J}^{-T} + {}^tg_{13}\ {}^0\mathbf{J}^{-1}\begin{bmatrix}0&0&1\\0&0&0\\1&0&0\end{bmatrix}{}^0\mathbf{J}^{-T} + {}^tg_{23}\ {}^0\mathbf{J}^{-1}\begin{bmatrix}0&0&0\\0&0&1\\0&1&0\end{bmatrix}{}^0\mathbf{J}^{-T}\end{aligned} \tag{16}$$

where ${}^tg_{ij}$ is defined to denote the component at $i$th row and $j$th column of the matrix ${}^t\mathbf{J}\ {}^t\mathbf{J}^T$ which is expressed as

$${}^t\mathbf{J}\ {}^t\mathbf{J}^T = \begin{bmatrix}{}^tg_{11} & {}^tg_{12} & {}^tg_{13} \\ {}^tg_{21} & {}^tg_{22} & {}^tg_{23} \\ {}^tg_{31} & {}^tg_{32} & {}^tg_{33}\end{bmatrix} \tag{17}$$

By recognising the tensor's symmetry, we define a six-component vector ${}^t\hat{\mathbf{g}}$ to be

$${}^t\hat{\mathbf{g}} = \begin{bmatrix}{}^tg_{11} & {}^tg_{22} & {}^tg_{33} & {}^tg_{12} & {}^tg_{13} & {}^tg_{23}\end{bmatrix}^T \tag{18}$$

Also, by recognising the constant matrix followed by each ${}^tg_{ij}$ component, ${}_0^t\mathbf{C}$ can be further written as

$${}_0^t\mathbf{C} = {}^tg_{11}\mathbf{G}_{11} + {}^tg_{22}\mathbf{G}_{22} + {}^tg_{33}\mathbf{G}_{33} + {}^tg_{12}\mathbf{G}_{12} + {}^tg_{13}\mathbf{G}_{13} + {}^tg_{23}\mathbf{G}_{23} \tag{19}$$

where $\mathbf{G}_{11}, \mathbf{G}_{22}, \mathbf{G}_{33}, \mathbf{G}_{12}, \mathbf{G}_{13}$ and $\mathbf{G}_{23}$ are defined as constant matrices (see Appendix B); the above proposed new expression of ${}_0^t\mathbf{C}$ is used to establish new expressions for the strain invariants and pseudo-invariants.

### 2.3.1 Formulation for isotropic hyperelasticity

For an incompressible isotropic hyperelastic material, the strain energy density function $\Psi$ depends on the three principal invariants $I_1$, $I_2$ and $I_3$ of ${}_0^t\mathbf{C}$ and is expressed by $\Psi({}_0^t\mathbf{C}) = \Psi(I_1, I_2, I_3)$ where $I_1$, $I_2$ and $I_3$ are given by

$$I_1 = \operatorname{tr}({}_0^t\mathbf{C}) \tag{20a}$$

$$I_2 = \frac{1}{2}\left[\left(\operatorname{tr}({}_0^t\mathbf{C})\right)^2 - \operatorname{tr}({}_0^t\mathbf{C}^2)\right] \tag{20b}$$

$$I_3 = \det({}_0^t\mathbf{C}) \tag{20c}$$

where $\operatorname{tr}(\cdot)$ denotes the trace of a matrix; if $\Psi({}_0^t\mathbf{C}) = \Psi(I_1, I_2, I_3)$ is expressed in the uncoupled form $\Psi({}_0^t\mathbf{C}) = \Psi(\bar{I}_1, \bar{I}_2) + \Psi(J)$, then $\bar{I}_1$ and $\bar{I}_2$ are called the modified strain invariants of $I_1$ and $I_2$, and they are related by $\bar{I}_1 = J^{-2/3}I_1$ and $\bar{I}_2 = J^{-4/3}I_2$.

The new formulations for $\operatorname{tr}({}_0^t\mathbf{C})$, $\operatorname{tr}({}_0^t\mathbf{C}^2)$ and $\det({}_0^t\mathbf{C})$ are first established in order to express $I_1$, $I_2$ and $I_3$.

- Based on the proposed new expression of ${}_0^t\mathbf{C}$ in Eq. (19), $\operatorname{tr}({}_0^t\mathbf{C})$ can be expressed as

$$\operatorname{tr}({}_0^t\mathbf{C}) = {}^tg_{11}\operatorname{tr}(\mathbf{G}_{11}) + {}^tg_{22}\operatorname{tr}(\mathbf{G}_{22}) + {}^tg_{33}\operatorname{tr}(\mathbf{G}_{33}) + {}^tg_{12}\operatorname{tr}(\mathbf{G}_{12}) + {}^tg_{13}\operatorname{tr}(\mathbf{G}_{13}) + {}^tg_{23}\operatorname{tr}(\mathbf{G}_{23}) \tag{21}$$

which can be written as

$$\operatorname{tr}({}_0^t\mathbf{C}) = {}^t\hat{\mathbf{g}}^T \mathbf{m}_1 \tag{22}$$

where ${}^t\hat{\mathbf{g}}$ is given in Eq. (18), and $\mathbf{m}_1$ is defined as a constant time-invariant six-component vector consisted of the traces of $\mathbf{G}$ matrices, and it can be precomputed. $\mathbf{m}_1$ is defined as

$$\mathbf{m}_1 = [\operatorname{tr}(\mathbf{G}_{11})\ \operatorname{tr}(\mathbf{G}_{22})\ \operatorname{tr}(\mathbf{G}_{33})\ \operatorname{tr}(\mathbf{G}_{12})\ \operatorname{tr}(\mathbf{G}_{13})\ \operatorname{tr}(\mathbf{G}_{23})]^T \tag{23}$$

- Similarly, $\operatorname{tr}({}_0^t\mathbf{C}^2)$ can be expressed as

$$\operatorname{tr}({}_0^t\mathbf{C}^2) = {}^t\hat{\mathbf{g}}^T \mathbf{W}\, {}^t\hat{\mathbf{g}} \tag{24}$$

where $\mathbf{W}$ is defined as a constant time-invariant matrix consisted of the traces of $\mathbf{G}$ matrices, and it can be precomputed. $\mathbf{W}$ is defined as

$$\mathbf{W} = \begin{bmatrix} \operatorname{tr}(\mathbf{G}_{11}{}^T\mathbf{G}_{11}) & \operatorname{tr}(\mathbf{G}_{11}{}^T\mathbf{G}_{22}) & \cdots & \operatorname{tr}(\mathbf{G}_{11}{}^T\mathbf{G}_{23}) \\ \operatorname{tr}(\mathbf{G}_{22}{}^T\mathbf{G}_{11}) & \operatorname{tr}(\mathbf{G}_{22}{}^T\mathbf{G}_{22}) & \cdots & \vdots \\ \vdots & \vdots & \ddots & \vdots \\ \operatorname{tr}(\mathbf{G}_{23}{}^T\mathbf{G}_{11}) & \cdots & \cdots & \operatorname{tr}(\mathbf{G}_{23}{}^T\mathbf{G}_{23}) \end{bmatrix} \tag{25}$$

- $\det({}_0^t\mathbf{C})$ can be expressed as

$$\det({}_0^t\mathbf{C}) = J^2 = \left(\det({}^t\mathbf{J})\det({}^0\mathbf{J})^{-1}\right)^2 \tag{26}$$

Using the above expressions, the new Jacobian formulations for $I_1$, $I_2$ and $I_3$ can be established.

- $I_1$ can be expressed as

$$I_1 = \operatorname{tr}({}_0^t\mathbf{C}) = {}^t\hat{\mathbf{g}}^T \mathbf{m}_1 \tag{27}$$

- $I_2$ can be expressed as

$$I_2 = \frac{1}{2}\left[(\mathrm{tr}(_0^t\mathbf{C}))^2 - \mathrm{tr}(_0^t\mathbf{C}^2)\right] = \frac{1}{2}\,{}^t\hat{\mathbf{g}}^T[\mathbf{m}_1\mathbf{m}_1^T - \mathbf{W}]\,{}^t\hat{\mathbf{g}} = {}^t\hat{\mathbf{g}}^T\mathbf{M}_2\,{}^t\hat{\mathbf{g}} \tag{28}$$

where $\mathbf{M}_2$ is defined as a constant time-invariant matrix $\mathbf{M}_2 = \frac{1}{2}[\mathbf{m}_1\mathbf{m}_1^T - \mathbf{W}]$, and it can be precomputed.

- $I_3$ can be expressed as

$$I_3 = \det(_0^t\mathbf{C}) = \left(\det({}^t\mathbf{J})\det({}^0\mathbf{J})^{-1}\right)^2 \tag{29}$$

### 2.3.2  Formulation for anisotropic hyperelasticity

For a fibre-reinforced composite incompressible hyperelastic material with two families of fibres in the orthotropic directions, $\Psi$ is expressed as $\Psi(_0^t\mathbf{C}) = \Psi(I_1, I_2, I_3, I_4, I_5, I_6, I_7)$ where $I_4$, $I_5$, $I_6$ and $I_7$ are called the pseudo-invariants which arise from the fibre-induced material anisotropy [29]. If the material is reinforced by only one family of fibres with a single preferred direction, it is said to be transversely isotropic with respect to this direction, and $\Psi$ is expressed as $\Psi(_0^t\mathbf{C}) = \Psi(I_1, I_2, I_3, I_4, I_5)$.

The pseudo-invariants $I_4$, $I_5$, $I_6$ and $I_7$ are given by

$$I_4 = \mathrm{tr}(\mathbf{A}_0^t\mathbf{C}) \tag{30a}$$

$$I_5 = \mathrm{tr}(\mathbf{A}_0^t\mathbf{C}^2) \tag{30b}$$

$$I_6 = \mathrm{tr}(\mathbf{B}_0^t\mathbf{C}) \tag{30c}$$

$$I_7 = \mathrm{tr}(\mathbf{B}_0^t\mathbf{C}^2) \tag{30d}$$

where the structural, fibre direction matrices $\mathbf{A} = {}^0\mathbf{a}\,{}^0\mathbf{a}^T$ and $\mathbf{B} = {}^0\mathbf{b}\,{}^0\mathbf{b}^T$ in which ${}^0\mathbf{a} = \begin{bmatrix} {}^0a_x & {}^0a_y & {}^0a_z \end{bmatrix}^T$ and ${}^0\mathbf{b} = \begin{bmatrix} {}^0b_x & {}^0b_y & {}^0b_z \end{bmatrix}^T$ are the unit vectors indicating the preferred fibre directions.

If expressing $\Psi$ in the uncoupled form $\Psi(_0^t\mathbf{C}) = \Psi(\bar{I}_1, \bar{I}_2, \bar{I}_4, \bar{I}_5, \bar{I}_6, \bar{I}_7) + \Psi(J)$, then $\bar{I}_4$, $\bar{I}_5$, $\bar{I}_6$ and $\bar{I}_7$ are called the modified pseudo-invariants of $I_4$, $I_5$, $I_6$ and $I_7$, and they are related by $\bar{I}_4 = J^{-2/3}I_4$, $\bar{I}_5 = J^{-4/3}I_5$, $\bar{I}_6 = J^{-2/3}I_6$ and $\bar{I}_7 = J^{-4/3}I_7$.

- Similar to $I_1 = \mathrm{tr}(_0^t\mathbf{C}) = {}^t\hat{\mathbf{g}}^T\mathbf{m}_1$ in Eq. (27), the pseudo-invariants $I_4$ and $I_6$ can be expressed as

$$I_4 = \mathrm{tr}(\mathbf{A}_0^t\mathbf{C}) = {}^t\hat{\mathbf{g}}^T\mathbf{m}_4 \tag{31}$$

$$I_6 = \mathrm{tr}(\mathbf{B}_0^t\mathbf{C}) = {}^t\hat{\mathbf{g}}^T\mathbf{m}_6 \tag{32}$$

where $\mathbf{m}_4$ and $\mathbf{m}_6$ are defined in a similar manner as $\mathbf{m}_1$ (see Appendix C), which are constant time-invariant vectors consisted of the traces of $\mathbf{AG}$ and $\mathbf{BG}$ matrices, respectively, and they can be precomputed.

- Similar to $\mathrm{tr}(_0^t\mathbf{C}^2) = {}^t\hat{\mathbf{g}}^T\mathbf{W}\,{}^t\hat{\mathbf{g}}$ in Eq. (24), the pseudo-invariants $I_5$ and $I_7$ can be expressed as

$$I_5 = \mathrm{tr}(\mathbf{A}_0^t\mathbf{C}^2) = {}^t\hat{\mathbf{g}}^T\mathbf{M}_5\,{}^t\hat{\mathbf{g}} \tag{33}$$

$$I_7 = \mathrm{tr}(\mathbf{B}_0^t\mathbf{C}^2) = {}^t\hat{\mathbf{g}}^T\mathbf{M}_7\,{}^t\hat{\mathbf{g}} \tag{34}$$

where $\mathbf{M}_5$ and $\mathbf{M}_7$ are defined in a similar way as $\mathbf{W}$ (see Appendix C), which are constant time-invariant matrices consisted of the traces of $\mathbf{AG}$ and $\mathbf{BG}$ matrices, respectively, and they can also be precomputed.

Table. 1 presents a summary of formulations for the principal strain invariants and pseudo-invariants in the proposed DJ-TLED.

**Table. 1** Formulations for the strain invariants and pseudo-invariants in the proposed DJ-TLED.

| [1]Invariants | Conventional formulations | [2]Proposed formulations | Equations |
|---|---|---|---|
| $I_1$ | $\mathrm{tr}(^t_0\mathbf{C})$ | $^t\hat{\mathbf{g}}^T \mathbf{m}_1$ | (27) |
| $I_2$ | $\frac{1}{2}\left[(\mathrm{tr}(^t_0\mathbf{C}))^2 - \mathrm{tr}(^t_0\mathbf{C}^2)\right]$ | $^t\hat{\mathbf{g}}^T \mathbf{M}_2\, ^t\hat{\mathbf{g}}$ | (28) |
| $I_3$ | $\det(^t_0\mathbf{C})$ | $\left(\det(^t\mathbf{J})\det(^0\mathbf{J})^{-1}\right)^2$ | (29) |
| $I_4$ | $\mathrm{tr}(\mathbf{A}^t_0\mathbf{C})$ | $^t\hat{\mathbf{g}}^T \mathbf{m}_4$ | (31) |
| $I_5$ | $\mathrm{tr}(\mathbf{A}^t_0\mathbf{C}^2)$ | $^t\hat{\mathbf{g}}^T \mathbf{M}_5\, ^t\hat{\mathbf{g}}$ | (33) |
| $I_6$ | $\mathrm{tr}(\mathbf{B}^t_0\mathbf{C})$ | $^t\hat{\mathbf{g}}^T \mathbf{m}_6$ | (32) |
| $I_7$ | $\mathrm{tr}(\mathbf{B}^t_0\mathbf{C}^2)$ | $^t\hat{\mathbf{g}}^T \mathbf{M}_7\, ^t\hat{\mathbf{g}}$ | (34) |

Note[1]: $\bar{I}_1 = J^{-2/3}I_1$, $\bar{I}_2 = J^{-4/3}I_2$, $\bar{I}_4 = J^{-2/3}I_4$, $\bar{I}_5 = J^{-4/3}I_5$, $\bar{I}_6 = J^{-2/3}I_6$ and $\bar{I}_7 = J^{-4/3}I_7$.
Note[2]: $\mathbf{m}_1$, $\mathbf{m}_4$ and $\mathbf{m}_6$ are constant time-invariant vectors, and $\mathbf{M}_2$, $\mathbf{M}_5$ and $\mathbf{M}_7$ are constant time-invariant matrices (see Appendix C). They can be precomputed.

*2.4 Proposed new time-invariant tensors $\mathbf{I}_{im}$ ($i = 1, 2, 4, 5, 6, 7$)*

As shown in Eq. (10), the proposed DJ-TLED formulation leads to a new constant time-invariant tensor $\mathbf{I}_{1m}$, which arises from the multiplication of constant components measured with respect to the reference (undeformed) configuration, and it can be precomputed. This section presents the formulations for the new time-invariant tensors $\mathbf{I}_{1m}$, $\mathbf{I}_{2m}$, $\mathbf{I}_{4m}$, $\mathbf{I}_{5m}$, $\mathbf{I}_{6m}$ and $\mathbf{I}_{7m}$.

*2.4.1 Formulation for isotropic hyperelasticity*

Recall the strain energy density function $\Psi(^t_0\mathbf{C}) = \Psi(\bar{I}_1) + \Psi(J)$, being dependent on $\bar{I}_1$ of $^t_0\bar{\mathbf{C}}$, the stress component $2J^{-2/3}\frac{\partial \Psi}{\partial \bar{I}_1}\mathbf{I}$ in the second Piola-Kirchhoff stress $^t_0\mathbf{S}$ in Eq. (8) is due to

$$2\frac{\partial \Psi}{\partial \bar{I}_1}\frac{\partial \bar{I}_1}{\partial ^t_0\mathbf{C}} = 2J^{-2/3}\frac{\partial \Psi}{\partial \bar{I}_1}\mathbf{I} \tag{35}$$

which leads to the following force contributions in the proposed direct Jacobian formulation:

$$^t_0\mathbf{X}\left(2J^{-2/3}\frac{\partial \Psi}{\partial \bar{I}_1}\mathbf{I}\right)\, ^0\mathbf{B}\, ^0V = \,^t\mathbf{J}^T\, ^0\mathbf{J}^{-T}\left(2J^{-2/3}\frac{\partial \Psi}{\partial \bar{I}_1}\mathbf{I}\right)\, ^0\mathbf{J}^{-1}\mathbf{H}_\xi\, ^0V = J^{-2/3}\, ^t\mathbf{J}^T\frac{\partial \Psi}{\partial \bar{I}_1}\mathbf{I}_{1m}\mathbf{H}_\xi \tag{36}$$

where $\mathbf{I}_{1m}$ is a constant time-invariant matrix defined as

$$\mathbf{I}_{1m} = 2\, ^0V\, ^0\mathbf{J}^{-T}\, ^0\mathbf{J}^{-1} \tag{37}$$

Similarly, if $\Psi$ is also dependent on $\bar{I}_2$ of $^t_0\bar{\mathbf{C}}$, i.e., $\Psi(^t_0\mathbf{C}) = \Psi(\bar{I}_1, \bar{I}_2) + \Psi(J)$, it leads to the following stress component

$$2\frac{\partial \Psi}{\partial \bar{I}_2}\frac{\partial \bar{I}_2}{\partial ^t_0\mathbf{C}} = 2J^{-4/3}\frac{\partial \Psi}{\partial \bar{I}_2}(I_1\mathbf{I} - ^t_0\mathbf{C}) \tag{38}$$

which leads to the following force contributions:

$$ {}_0^t\mathbf{X}\left(2J^{-4/3}\frac{\partial \Psi}{\partial \bar{I}_2}(I_1\mathbf{I} - {}_0^t\mathbf{C})\right) {}^0\mathbf{B}\,{}^0V = J^{-4/3}\,{}^t\mathbf{J}^T\frac{\partial \Psi}{\partial \bar{I}_2}\left({}^t\hat{\mathbf{g}}^T\cdot \mathbf{I}_{2m}\right)\mathbf{H}_\xi \qquad (39) $$

where $\mathbf{I}_{2m}$ consists of six constant matrices, and it is defined as

$$ \mathbf{I}_{2m} = \begin{bmatrix} \mathbf{I}_{2,11} & \mathbf{I}_{2,22} & \mathbf{I}_{2,33} & \mathbf{I}_{2,12} & \mathbf{I}_{2,13} & \mathbf{I}_{2,23} \end{bmatrix} \qquad (40) $$

where $\mathbf{I}_{2,11}, \mathbf{I}_{2,22}, \mathbf{I}_{2,33}, \mathbf{I}_{2,12}, \mathbf{I}_{2,13}$ and $\mathbf{I}_{2,23}$ are defined as

$$ \mathbf{I}_{2,11} = 2\,{}^0V\,{}^0\mathbf{J}^{-T}(\mathrm{tr}(\mathbf{G}_{11})\mathbf{I} - \mathbf{G}_{11})\,{}^0\mathbf{J}^{-1} \qquad (41\text{a}) $$

$$ \mathbf{I}_{2,22} = 2\,{}^0V\,{}^0\mathbf{J}^{-T}(\mathrm{tr}(\mathbf{G}_{22})\mathbf{I} - \mathbf{G}_{22})\,{}^0\mathbf{J}^{-1} \qquad (41\text{b}) $$

$$ \mathbf{I}_{2,33} = 2\,{}^0V\,{}^0\mathbf{J}^{-T}(\mathrm{tr}(\mathbf{G}_{33})\mathbf{I} - \mathbf{G}_{33})\,{}^0\mathbf{J}^{-1} \qquad (41\text{c}) $$

$$ \mathbf{I}_{2,12} = 2\,{}^0V\,{}^0\mathbf{J}^{-T}(\mathrm{tr}(\mathbf{G}_{12})\mathbf{I} - \mathbf{G}_{12})\,{}^0\mathbf{J}^{-1} \qquad (41\text{d}) $$

$$ \mathbf{I}_{2,13} = 2\,{}^0V\,{}^0\mathbf{J}^{-T}(\mathrm{tr}(\mathbf{G}_{13})\mathbf{I} - \mathbf{G}_{13})\,{}^0\mathbf{J}^{-1} \qquad (41\text{e}) $$

$$ \mathbf{I}_{2,23} = 2\,{}^0V\,{}^0\mathbf{J}^{-T}(\mathrm{tr}(\mathbf{G}_{23})\mathbf{I} - \mathbf{G}_{23})\,{}^0\mathbf{J}^{-1} \qquad (41\text{f}) $$

In Eq. (39), $\left({}^t\hat{\mathbf{g}}^T\cdot\mathbf{I}_{2m}\right)$ is computed in the following manner

$$ {}^t\hat{\mathbf{g}}^T\cdot\mathbf{I}_{2m} = {}^tg_{11}\mathbf{I}_{2,11} + {}^tg_{22}\mathbf{I}_{2,22} + {}^tg_{33}\mathbf{I}_{2,33} + {}^tg_{12}\mathbf{I}_{2,12} + {}^tg_{13}\mathbf{I}_{2,13} + {}^tg_{23}\mathbf{I}_{2,23} \qquad (42) $$

### 2.4.2 Formulation for anisotropic hyperelasticity

Consider the strain energy density function $\Psi({}_0^t\mathbf{C}) = \Psi(\bar{I}_1, \bar{I}_2, \bar{I}_4, \bar{I}_5, \bar{I}_6, \bar{I}_7) + \Psi(J)$ for anisotropic hyperelasticity, it leads to the following stress component related to $\bar{I}_4$:

$$ 2\frac{\partial \Psi}{\partial \bar{I}_4}\frac{\partial \bar{I}_4}{\partial {}_0^t\mathbf{C}} = 2J^{-2/3}\frac{\partial \Psi}{\partial \bar{I}_4}\mathbf{A} \qquad (43) $$

which leads to the following force contributions in the proposed direct Jacobian formulation:

$$ {}_0^t\mathbf{X}\left(2J^{-2/3}\frac{\partial \Psi}{\partial \bar{I}_4}\mathbf{A}\right){}^0\mathbf{B}\,{}^0V = J^{-2/3}\,{}^t\mathbf{J}^T\frac{\partial \Psi}{\partial \bar{I}_4}\mathbf{I}_{4m}\mathbf{H}_\xi \qquad (44) $$

where $\mathbf{I}_{4m}$ (similar to the expression of $\mathbf{I}_{1m}$) is a constant time-invariant matrix concerning the fibre direction matrix $\mathbf{A}$, and it is defined as

$$ \mathbf{I}_{4m} = 2\,{}^0V\,{}^0\mathbf{J}^{-T}\mathbf{A}\,{}^0\mathbf{J}^{-1} \qquad (45) $$

Similarly, $\mathbf{I}_{6m}$ concerning the fibre direction matrix $\mathbf{B}$ is defined as

$$ \mathbf{I}_{6m} = 2\,{}^0V\,{}^0\mathbf{J}^{-T}\mathbf{B}\,{}^0\mathbf{J}^{-1} \qquad (46) $$

We now define $\mathbf{I}_{5m}$ and $\mathbf{I}_{7m}$. The strain energy density function $\Psi({}_0^t\mathbf{C})$ leads to the following stress component related to $\bar{I}_5$:

$$ 2\frac{\partial \Psi}{\partial \bar{I}_5}\frac{\partial \bar{I}_5}{\partial {}_0^t\mathbf{C}} = 2J^{-4/3}\frac{\partial \Psi}{\partial \bar{I}_5}(\mathbf{A}\,{}_0^t\mathbf{C} + {}_0^t\mathbf{C}\mathbf{A}) \qquad (47) $$

which leads to the following force contributions:

$$
{}_0^t\mathbf{X}\left(2J^{-4/3}\frac{\partial\Psi}{\partial\bar{I}_5}(\mathbf{A}{}_0^t\mathbf{C}+{}_0^t\mathbf{C}\mathbf{A})\right){}^0\mathbf{B}\,{}^0V = J^{-4/3}\,{}^t\mathbf{J}^T\frac{\partial\Psi}{\partial\bar{I}_5}\left({}^t\hat{\mathbf{g}}^T\cdot\mathbf{I}_{5m}\right)\mathbf{H}_\xi
\tag{48}
$$

where $\mathbf{I}_{5m}$ (similar to the expression of $\mathbf{I}_{2m}$) consists of six constant matrices concerning the fibre direction matrix $\mathbf{A}$, and it is defined as

$$
\mathbf{I}_{5m} = \begin{bmatrix}\mathbf{I}_{5,11}\ \mathbf{I}_{5,22}\ \mathbf{I}_{5,33}\ \mathbf{I}_{5,12}\ \mathbf{I}_{5,13}\ \mathbf{I}_{5,23}\end{bmatrix}
\tag{49}
$$

where $\mathbf{I}_{5,11}, \mathbf{I}_{5,22}, \mathbf{I}_{5,33}, \mathbf{I}_{5,12}, \mathbf{I}_{5,13}$ and $\mathbf{I}_{5,23}$ are defined as

$$
\mathbf{I}_{5,11} = 2\,{}^0V\,{}^0\mathbf{J}^{-T}(\mathbf{A}\mathbf{G}_{11}+\mathbf{G}_{11}\mathbf{A})\,{}^0\mathbf{J}^{-1}
\tag{50a}
$$

$$
\mathbf{I}_{5,22} = 2\,{}^0V\,{}^0\mathbf{J}^{-T}(\mathbf{A}\mathbf{G}_{22}+\mathbf{G}_{22}\mathbf{A})\,{}^0\mathbf{J}^{-1}
\tag{50b}
$$

$$
\mathbf{I}_{5,33} = 2\,{}^0V\,{}^0\mathbf{J}^{-T}(\mathbf{A}\mathbf{G}_{33}+\mathbf{G}_{33}\mathbf{A})\,{}^0\mathbf{J}^{-1}
\tag{50c}
$$

$$
\mathbf{I}_{5,12} = 2\,{}^0V\,{}^0\mathbf{J}^{-T}(\mathbf{A}\mathbf{G}_{12}+\mathbf{G}_{12}\mathbf{A})\,{}^0\mathbf{J}^{-1}
\tag{50d}
$$

$$
\mathbf{I}_{5,13} = 2\,{}^0V\,{}^0\mathbf{J}^{-T}(\mathbf{A}\mathbf{G}_{13}+\mathbf{G}_{13}\mathbf{A})\,{}^0\mathbf{J}^{-1}
\tag{50e}
$$

$$
\mathbf{I}_{5,23} = 2\,{}^0V\,{}^0\mathbf{J}^{-T}(\mathbf{A}\mathbf{G}_{23}+\mathbf{G}_{23}\mathbf{A})\,{}^0\mathbf{J}^{-1}
\tag{50f}
$$

Similarly, $\mathbf{I}_{7m}$ concerning the fibre direction matrix $\mathbf{B}$ is defined as

$$
\mathbf{I}_{7m} = \begin{bmatrix}\mathbf{I}_{7,11}\ \mathbf{I}_{7,22}\ \mathbf{I}_{7,33}\ \mathbf{I}_{7,12}\ \mathbf{I}_{7,13}\ \mathbf{I}_{7,23}\end{bmatrix}
\tag{51}
$$

where $\mathbf{I}_{7,11}, \mathbf{I}_{7,22}, \mathbf{I}_{7,33}, \mathbf{I}_{7,12}, \mathbf{I}_{7,13}$ and $\mathbf{I}_{7,23}$ are defined in a similar way as those of $\mathbf{I}_{5m}$ by substituting $\mathbf{B}$ for $\mathbf{A}$.

Table. 2 presents a summary of formulations of $\mathbf{I}_{1m}, \mathbf{I}_{2m}, \mathbf{I}_{4m}, \mathbf{I}_{5m}, \mathbf{I}_{6m}$ and $\mathbf{I}_{7m}$ in the proposed DJ-TLED.

**Table. 2** Formulations of $\mathbf{I}_{1m}, \mathbf{I}_{2m}, \mathbf{I}_{4m}, \mathbf{I}_{5m}, \mathbf{I}_{6m}$ and $\mathbf{I}_{7m}$ in the proposed DJ-TLED.

|   | Proposed formulations | Equations |
|---|---|---|
| $\mathbf{I}_{1m}$ | $2\,{}^0V\,{}^0\mathbf{J}^{-T}\,{}^0\mathbf{J}^{-1}$ | (37) |
| $\mathbf{I}_{2m}$ | $[\mathbf{I}_{2,11}\ \mathbf{I}_{2,22}\ \mathbf{I}_{2,33}\ \mathbf{I}_{2,12}\ \mathbf{I}_{2,13}\ \mathbf{I}_{2,23}]$ | (40) |
| $\mathbf{I}_{4m}$ | $2\,{}^0V\,{}^0\mathbf{J}^{-T}\mathbf{A}\,{}^0\mathbf{J}^{-1}$ | (45) |
| $\mathbf{I}_{5m}$ | $[\mathbf{I}_{5,11}\ \mathbf{I}_{5,22}\ \mathbf{I}_{5,33}\ \mathbf{I}_{5,12}\ \mathbf{I}_{5,13}\ \mathbf{I}_{5,23}]$ | (49) |
| $\mathbf{I}_{6m}$ | $2\,{}^0V\,{}^0\mathbf{J}^{-T}\mathbf{B}\,{}^0\mathbf{J}^{-1}$ | (46) |
| $\mathbf{I}_{7m}$ | $[\mathbf{I}_{7,11}\ \mathbf{I}_{7,22}\ \mathbf{I}_{7,33}\ \mathbf{I}_{7,12}\ \mathbf{I}_{7,13}\ \mathbf{I}_{7,23}]$ | (51) |

Note: $\mathbf{I}_{1m}, \mathbf{I}_{2m}, \mathbf{I}_{4m}, \mathbf{I}_{5m}, \mathbf{I}_{6m}$ and $\mathbf{I}_{7m}$ can be precomputed.

## 3. Examples of hyperelastic force formulations in DJ-TLED

The proposed DJ-TLED is established in a general way in terms of the strain energy density function to allow for any hyperelastic constitutive model to be incorporated. The procedure to derive the nodal force contributions for a strain energy density function in the proposed DJ-TLED can be summarised as follows:

Step #1: Expressing the second Piola-Kirchhoff stress tensor ${}_0^t\mathbf{S}$ based on the given strain energy density function $\Psi$, such as Eq. (8),

Step #2: Reformulating the nodal force contributions ${}_0^t\mathbf{F}$ in TLED by the new DJ-TLED, such as Eq. (10),

Step #3: Using the proposed formulations for strain invariants (see Table. 1) and constant $\mathbf{I}_{im}$ (see Table. 2) for DJ-TLED computation.

Table. 3 presents examples of nodal force contributions $^t_0\mathbf{F}$ for some hyperelastic material models in DJ-TLED. It is worth recalling that $\mathbf{I}_{1m}$, $\mathbf{I}_{2m}$, $\mathbf{I}_{4m}$, $\mathbf{I}_{6m}$ and $\mathbf{H}_\xi$ are constant tensors that do not require run-time updates. Together with constants $^0V$, $\mu$, $\kappa$, $\eta_\mathbf{a}$, $\eta_\mathbf{b}$, $C_{10}$ and $C_{01}$, they yield very efficient online computation.

**Table. 3** Examples of nodal force contributions in the proposed DJ-TLED.

| Some generic forms of nodal force contributions $^t_0\mathbf{F}$ |
|---|
| $\Psi(^t_0\mathbf{C}) = \Psi(\bar{I}_1) + \Psi(J)$ |
| $^t_0\mathbf{F} = \left( J^{-2/3} \, ^t\mathbf{J}^T \left( \frac{\partial \Psi}{\partial \bar{I}_1} \mathbf{I}_{1m} \right) + \left( -\frac{2}{3} \left( \frac{\partial \Psi}{\partial \bar{I}_1} \bar{I}_1 \right) + J \frac{\partial \Psi}{\partial J} \right) \, ^0V \, ^t\mathbf{J}^{-1} \right) \mathbf{H}_\xi$ |
| $\Psi(^t_0\mathbf{C}) = \Psi(\bar{I}_1, \bar{I}_4) + \Psi(J)$ |
| $^t_0\mathbf{F} = \left( J^{-2/3} \, ^t\mathbf{J}^T \left( \frac{\partial \Psi}{\partial \bar{I}_1} \mathbf{I}_{1m} + \frac{\partial \Psi}{\partial \bar{I}_4} \mathbf{I}_{4m} \right) + \left( -\frac{2}{3} \left( \frac{\partial \Psi}{\partial \bar{I}_1} \bar{I}_1 + \frac{\partial \Psi}{\partial \bar{I}_4} \bar{I}_4 \right) + J \frac{\partial \Psi}{\partial J} \right) \, ^0V \, ^t\mathbf{J}^{-1} \right) \mathbf{H}_\xi$ |
| $\Psi(^t_0\mathbf{C}) = \Psi(\bar{I}_1, \bar{I}_4, \bar{I}_6) + \Psi(J)$ |
| $^t_0\mathbf{F} = \left( J^{-2/3} \, ^t\mathbf{J}^T \left( \frac{\partial \Psi}{\partial \bar{I}_1} \mathbf{I}_{1m} + \frac{\partial \Psi}{\partial \bar{I}_4} \mathbf{I}_{4m} + \frac{\partial \Psi}{\partial \bar{I}_6} \mathbf{I}_{6m} \right) + \left( -\frac{2}{3} \left( \frac{\partial \Psi}{\partial \bar{I}_1} \bar{I}_1 + \frac{\partial \Psi}{\partial \bar{I}_4} \bar{I}_4 + \frac{\partial \Psi}{\partial \bar{I}_6} \bar{I}_6 \right) + J \frac{\partial \Psi}{\partial J} \right) \, ^0V \, ^t\mathbf{J}^{-1} \right) \mathbf{H}_\xi$ |
| $\Psi(^t_0\mathbf{C}) = \Psi(\bar{I}_1, \bar{I}_2) + \Psi(J)$ |
| $^t_0\mathbf{F} = \left( J^{-2/3} \, ^t\mathbf{J}^T \left( \frac{\partial \Psi}{\partial \bar{I}_1} \mathbf{I}_{1m} + J^{-2/3} \frac{\partial \Psi}{\partial \bar{I}_2} (\, ^t\hat{\mathbf{g}}^T \cdot \mathbf{I}_{2m}) \right) + \left( -\frac{2}{3} \left( \frac{\partial \Psi}{\partial \bar{I}_1} \bar{I}_1 + 2 \frac{\partial \Psi}{\partial \bar{I}_2} \bar{I}_2 \right) + J \frac{\partial \Psi}{\partial J} \right) \, ^0V \, ^t\mathbf{J}^{-1} \right) \mathbf{H}_\xi$ |
| Examples for some hyperelastic material models |
| Neo-Hookean (NH) model: $\quad \Psi = \frac{\mu}{2}(\bar{I}_1 - 3) + \frac{\kappa}{2}(J - 1)^2$ |
| $^t_0\mathbf{F} = \left( J^{-2/3} \, ^t\mathbf{J}^T \left( \frac{\mu}{2} \mathbf{I}_{1m} \right) + \left( -\frac{2}{3} \left( \frac{\mu}{2} \bar{I}_1 \right) + \kappa J(J - 1) \right) \, ^0V \, ^t\mathbf{J}^{-1} \right) \mathbf{H}_\xi$ |
| Transversely isotropic neo-Hookean (TI) model: $\quad \Psi = \frac{\mu}{2}(\bar{I}_1 - 3) + \frac{\eta_\mathbf{a}}{2}(\bar{I}_4 - 1)^2 + \frac{\kappa}{2}(J - 1)^2$ |
| $^t_0\mathbf{F} = \left( J^{-2/3} \, ^t\mathbf{J}^T \left( \frac{\mu}{2} \mathbf{I}_{1m} + \eta_\mathbf{a}(\bar{I}_4 - 1)\mathbf{I}_{4m} \right) + \left( -\frac{2}{3} \left( \frac{\mu}{2} \bar{I}_1 + \eta_\mathbf{a}(\bar{I}_4 - 1)\bar{I}_4 \right) + \kappa J(J - 1) \right) \, ^0V \, ^t\mathbf{J}^{-1} \right) \mathbf{H}_\xi$ |
| Orthotropic neo-Hookean (OT) model: $\quad \Psi = \frac{\mu}{2}(\bar{I}_1 - 3) + \frac{\eta_\mathbf{a}}{2}(\bar{I}_4 - 1)^2 + \frac{\eta_\mathbf{b}}{2}(\bar{I}_6 - 1)^2 + \frac{\kappa}{2}(J - 1)^2$ |
| $^t_0\mathbf{F} = \left( J^{-2/3} \, ^t\mathbf{J}^T \left( \frac{\mu}{2} \mathbf{I}_{1m} + \eta_\mathbf{a}(\bar{I}_4 - 1)\mathbf{I}_{4m} + \eta_\mathbf{b}(\bar{I}_6 - 1)\mathbf{I}_{6m} \right) + \left( -\frac{2}{3} \left( \frac{\mu}{2} \bar{I}_1 + \eta_\mathbf{a}(\bar{I}_4 - 1)\bar{I}_4 + \eta_\mathbf{b}(\bar{I}_6 - 1)\bar{I}_6 \right) + \kappa J(J - 1) \right) \, ^0V \, ^t\mathbf{J}^{-1} \right) \mathbf{H}_\xi$ |
| Mooney-Rivlin (MR) model: $\quad \Psi = C_{10}(\bar{I}_1 - 3) + C_{01}(\bar{I}_2 - 3) + \frac{\kappa}{2}(J - 1)^2$ |
| $^t_0\mathbf{F} = \left( J^{-2/3} \, ^t\mathbf{J}^T \left( C_{10}\mathbf{I}_{1m} + J^{-2/3} C_{01}(\, ^t\hat{\mathbf{g}}^T \cdot \mathbf{I}_{2m}) \right) + \left( -\frac{2}{3} (C_{10}\bar{I}_1 + 2C_{01}\bar{I}_2) + \kappa J(J - 1) \right) \, ^0V \, ^t\mathbf{J}^{-1} \right) \mathbf{H}_\xi$ |
| Note: $\mu$ is the shear modulus, and $\kappa$ is the bulk modulus. |

## 4. GPU implementation

Having established the formulation for nodal force contributions, the proposed DJ-TLED is implemented using GPU acceleration for real-time computation of hyperelastic deformations. The GPU solution procedure consists of a host (CPU) and a device (GPU) implementation for host-device interaction.

The host implementation is written in the C++ programming language using Visual Studio 2017 for computing the constant time-invariant vectors $\mathbf{m}_1$, $\mathbf{m}_4$ and $\mathbf{m}_6$ and matrices $\mathbf{M}_2$, $\mathbf{M}_5$ and $\mathbf{M}_7$ (Appendix C) and $\mathbf{I}_{1m}$, $\mathbf{I}_{2m}$, $\mathbf{I}_{4m}$, $\mathbf{I}_{5m}$, $\mathbf{I}_{6m}$ and $\mathbf{I}_{7m}$ (Table. 2), all of which can be precomputed. The host implementation is also used to invoke methods on GPU for memory allocation, texture binding, data copy from/to device, and kernel launching.

The device implementation is written using the NVIDIA CUDA programming API for computing the new nodal internal forces and nodal displacements at each time step. The nodal force contributions are calculated by a kernel $k_f$ for element calculation. The new nodal displacements are computed by a kernel $k_u$ for node calculation. A time-stepping procedure is achieved by launching the kernel $k_f$ across $n_f$ threads and kernel $k_u$ across $n_u$ threads where $n_f$ and $n_u$ denote the number of elements and nodes in the model, respectively. GPU solution is achieved via a time-loop of such time-stepping procedures.

Algorithm. 1 presents the implementation of the proposed DJ-TLED. In CPU implementation, the "Loop over elements" is achieved using the *for*-loop in the C++ programming language. In GPU implementation, it is achieved by launching the kernel $k_f$ across $n_f$ threads. The time integration is performed using the central-difference scheme.

**Algorithm. 1** Implementation of the proposed DJ-TLED.

| |
|---|
| Pre-computation: |
|     Loop over elements: initialise/compute |
| 1:  $^0\mathbf{J}$, $\det(^0\mathbf{J})$, $^0V$, $\mathbf{A}$, $\mathbf{B}$ and $\mathbf{H}_\xi$. |
| 2:  $\mathbf{G}$ matrices (Appendix B). |
| 3:  $\mathbf{m}_1$, $\mathbf{m}_4$, $\mathbf{m}_6$, $\mathbf{M}_2$, $\mathbf{M}_5$ and $\mathbf{M}_7$ (Appendix C). |
| 4:  $\mathbf{I}_{1m}$, $\mathbf{I}_{2m}$, $\mathbf{I}_{4m}$, $\mathbf{I}_{5m}$, $\mathbf{I}_{6m}$ and $\mathbf{I}_{7m}$ (Table. 2). |
| Runtime-computation: |
|     Time stepping: |
|     Loop over elements: compute |
| 1:  $^t\mathbf{J}$, $^t\mathbf{J}^{-1}$ (Eq. (11)) and $J$ (Eq. (12)). |
| 2:  $^t\hat{\mathbf{g}}$ (Eq. (18)). |
| 3:  $I_1$, $I_2$, $I_4$, $I_5$, $I_6$ and $I_7$ (Table. 1). |
| 4:  $^t_0\mathbf{F}$ (Table. 3). |
|     Make a time step. |

## 5. Results

The proposed DJ-TLED is evaluated against the established TLED solution procedures for algorithm verification and CPU and GPU performance evaluation.

*5.1 Isotropic, transversely isotropic, and orthotropic hyperelastic deformations*

Fig. 2 illustrates the model employed for computation of isotropic and anisotropic hyperelastic deformations for algorithm verification. The model has a cubic shape ($0.1\ m \times 0.1\ m \times 0.1\ m$) with four circular cut-outs (height $= 0.03\ m$ with radius $= 0.05\ m$) in the centre of four faces. The back face of the model (without the circular cut-out) was assumed to be fixed in position, and the front (opposite) face was prescribed with an extension of $0.05\ m$ in the z-direction by a ramped displacement $u_z$ ( $^0u_z = 0$, $^tu_z = \frac{t}{t_{total}} * 0.05\ m$).

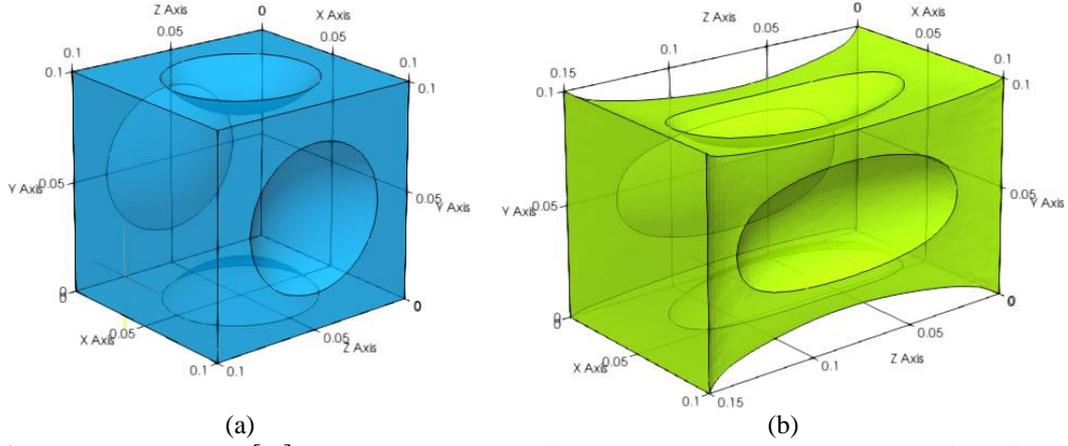

(a)                                       (b)

**Fig. 2.** (a) Model geometry $[m]$, and (b) an extension of $0.05\ m$ in the $z$-direction is applied for deformation.

The model was discretised into 4-node tetrahedral (T4) and 8-node hexahedral (H8) finite elements to simulate deformations of NH, TI, OT, and MR hyperelastic material models (see Table. 3). For NH, shear modulus $\mu = 6567\ Pa$ and bulk modulus $\kappa = 326210\ Pa$ (corresponding to Poisson's ratio $\nu = 0.49$) were employed. For TI, $\eta_a = 2\mu$ [30] was used with a preferred unit fibre direction $^0\mathbf{a} = [1\ 0\ 0]^T$ in the $x$-direction for material anisotropy. For OT, $\eta_a = 2\mu$ and $\eta_b = 2\mu$ were used with $^0\mathbf{a} = [1\ 0\ 0]^T$ in the $x$-direction and $^0\mathbf{b} = [0\ 1\ 0]^T$ in the $y$-direction for material anisotropy. For MR, $C_{10} = \frac{\mu}{2}$ (same as NH for $\bar{I}_1$) and $C_{01} = 3000\ Pa$ were used to observe the effect on $\bar{I}_2$ (corresponding to $\mu = 2(C_{01} + C_{10}) = 12567\ Pa$ and Poisson's ratio $\nu = 0.481$). It is worth noting that MR reduces to NH when $C_{01} = 0$, which is a special case of MR. Mass density $\rho = 1060\ kg/m^3$ was applied to all material models.

The model was discretised into 6 different mesh sizes ranged from 0.4 to $10.1 \times 10^4$ degrees of freedom (DOFs, 3 DOFs per node due to displacements in $x$, $y$ and $z$). The physical time to be simulated was $t = 1\ s$ with a time step size $\Delta t = 0.00005\ s$ (20000 steps) to ensure numerical stability of explicit dynamics at the largest model size ($10.1 \times 10^4$ DOFs). Due to using the low-order H8 finite elements with reduced integration (1-point Gauss) in the hexahedral mesh, an hourglass control algorithm [24] was applied to suppress the zero energy (hourglass) modes.

Fig. 3 presents the deformation results. Compared to the uniform contraction in NH, TI exhibited a pronounced stiffening in the $x$-direction due to the preferred fibre direction $^0\mathbf{a}$ and yielded more contraction in the $y$-direction. Likewise, OT exhibited stiffened deformations in both $x$- and $y$-directions near the edges of the model due to the fibre directions $^0\mathbf{a}$ and $^0\mathbf{b}$, it also experienced more contraction near the four circular cut-outs. On the other hand, MR exhibited fewer contraction near the circular cut-outs but had more deformations at the model edges, which could be due to the smaller Poisson's ratio (0.481 in MR compared to 0.49 in NH).

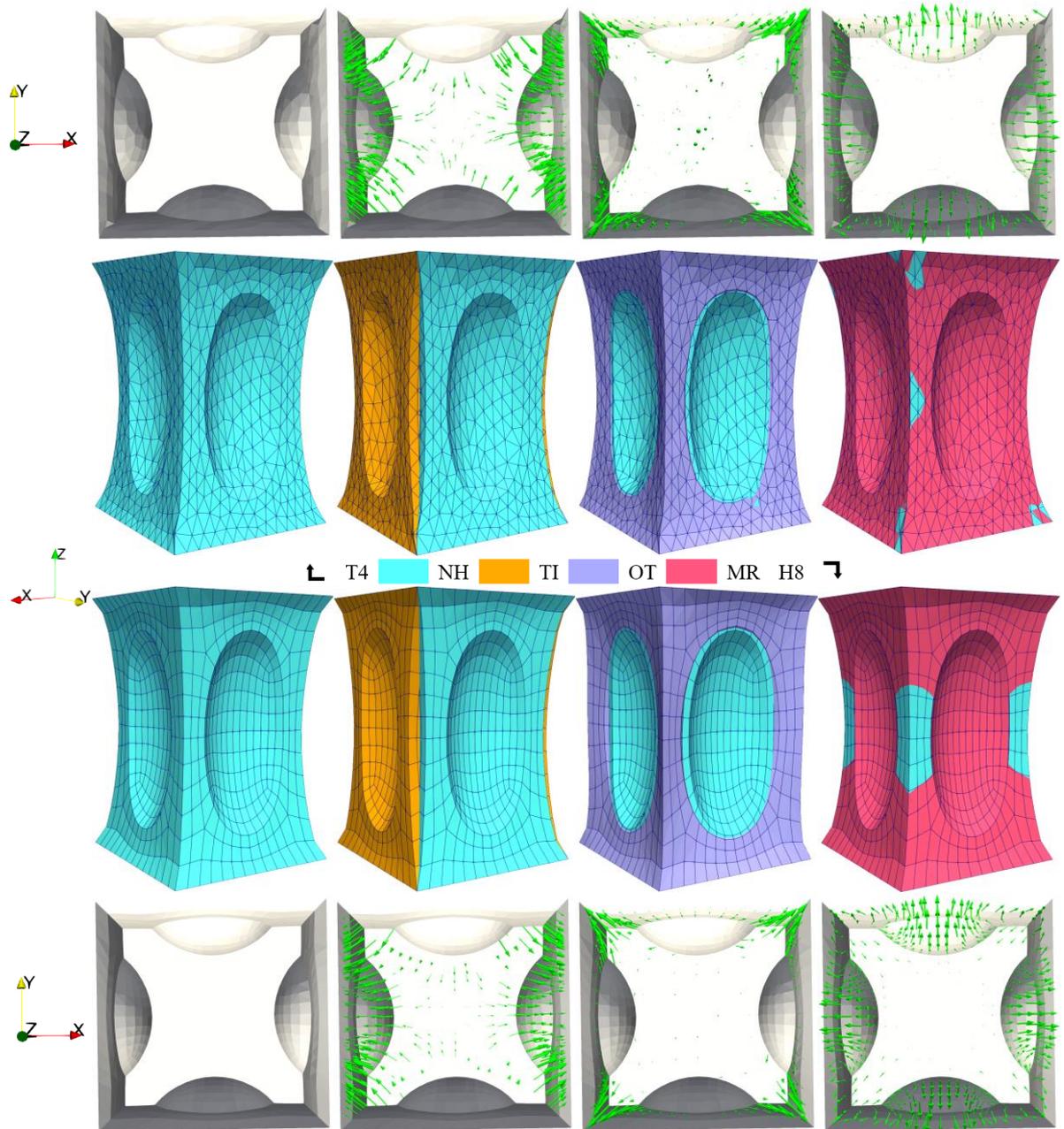

**Fig. 3.** Deformations of the neo-Hookean (NH), transversely isotropic neo-Hookean (TI), orthotropic neo-Hookean (OT), and Mooney-Rivlin (MR) hyperelastic constitutive models computed using the 4-node tetrahedral (T4) mesh and the 8-node hexahedral (H8) mesh by the proposed DJ-TLED (the arrows are normalised).

Fig. 4 presents the validation of numerical accuracy of the proposed DJ-TLED against TLED reference solutions for 6 different model sizes. As DJ-TLED uses the same fundamental basis (nonlinear continuum mechanics) as the standard TLED, except that DJ-TLED employs solely the Jacobian operator, both methods should intrinsically lead to the same level of numerical accuracy. In practice, however, some computation discrepancies may be observed between DJ-TLED and TLED due to using the different procedures for arithmetic floating-point operations. We used single precision floating-point numbers in C++ for cost-effective computation and less memory storage. In Fig. 4, the Root Mean Square Error (RMSE) decreased with the increase of model DOFs, and most RMSEs were within the 7 decimal digits of precision of the single precision floating-point. The RMSE between DJ-TLED and TLED is due to the small changes in single precision floating-point numbers during the different arithmetic operations (see Comparison with TLED in Section 7). The close match (small RMSE) between the DJ-TLED and TLED solutions demonstrates the validity of the proposed DJ-TLED for computation of isotropic and anisotropic hyperelastic deformations. RMSE was computed by

$$RMSE = \sqrt{\frac{\sum_{i=1}^{N}\left(u_i^{DJ-TLED} - u_i^{TLED}\right)^2}{N}} \qquad (52)$$

where $N$ is the number of DOFs.

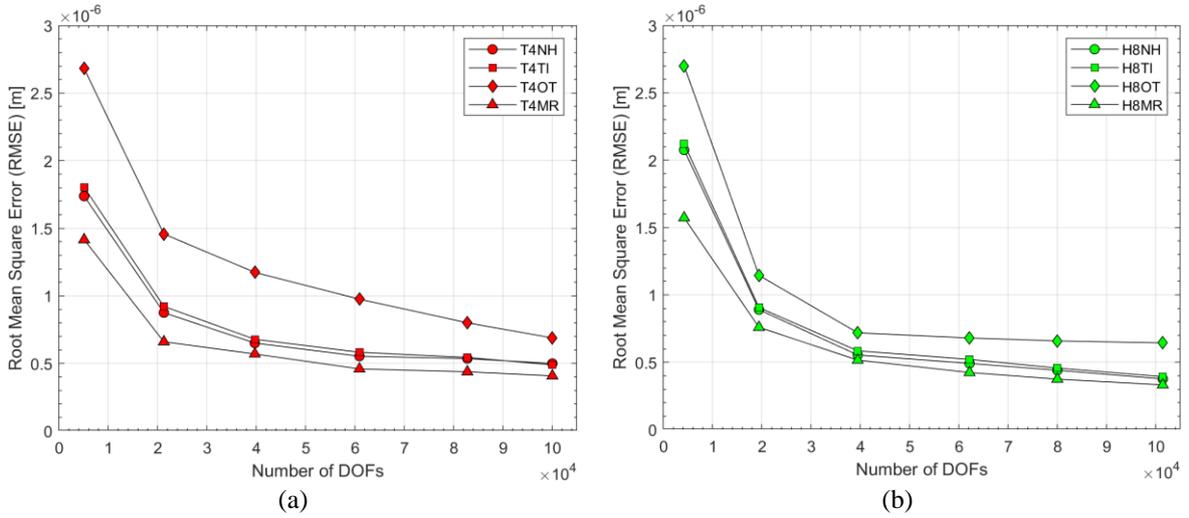

**Fig. 4.** Numerical accuracy at various model sizes in the (a) T4 mesh and (b) H8 mesh for NH, TI, OT, and MR material models.

*5.2 Large extension, compression, and shear deformations*

The model in Fig. 2 was also used to simulate large deformations of hyperelastic materials under extension, compression, and shear. As illustrated in Fig. 5, a ramped displacement $u_z = 0.2\ m$ was prescribed to induce extension, $u_z = -0.06\ m$ to induce compression, and $u_y = 0.2\ m$ to induce shear. These displacements were 200%, 60%, and 200% of the undeformed edge length (0.1 $m$) for large finite-strain deformations.

The deformations were computed using the NH material model with the same parameter values as those in Section 5.1. Furthermore, the established fully nonlinear finite element procedures with implicit time stepping, "Dynamic, Implicit", from commercial finite element analysis package, ABAQUS/CAE 2018 (2017_11_08-04.21.41 127140), were used with a smaller time step size $\Delta t = 0.000025\ s$ to produce reference solutions for validation. Table. 4 presents a comparison of RMSEs of DJ-TLED and TLED compared to ABAQUS references solutions. The close match (small RMSE) between the DJ-TLED and ABAQUS solutions demonstrates the validity of the proposed DJ-TLED for computation of large finite-strain deformations. Moreover, the RMSEs of DJ-TLED and TLED were almost identical, supporting the fact that both methods have the same level of numerical accuracy.

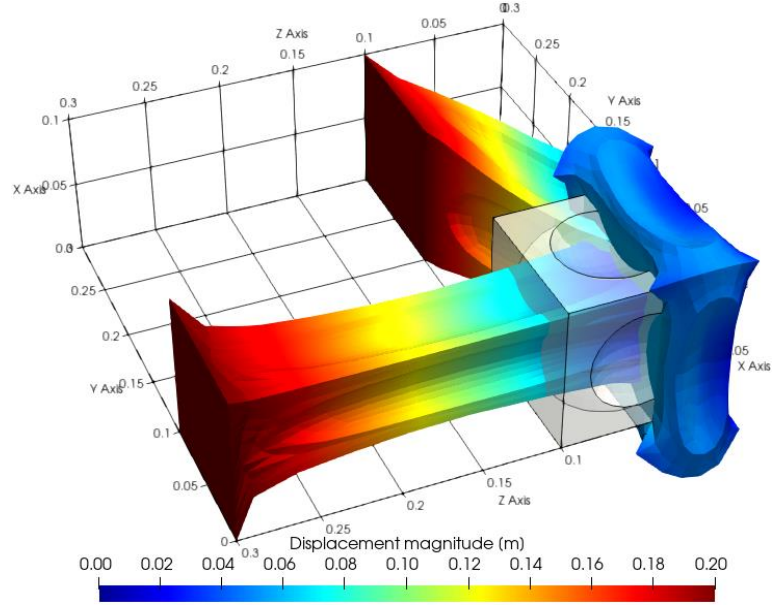

**Fig. 5.** The model is simulated using the NH material model for large deformations of extension, compression, and shear.

**Table. 4** Comparison of RMSEs between DJ-TLED and TLED against ABAQUS reference solutions.

|  |  | T4 | | H8 | |
|---|---|---|---|---|---|
|  |  | (1695 nodes, 7726 elements) | | (1398 nodes, 948 elements) | |
| RMSE | | DJ-TLED | TLED | DJ-TLED | TLED |
| (compared to ABAQUS) | E | 0.00024 | 0.00025 | 0.00118 | 0.00118 |
|  | C | 4.94080e-5 | 4.94968e-5 | 0.00341 | 0.00341 |
|  | S | 0.00052 | 0.00052 | 0.00129 | 0.00129 |
|  |  | | | E: extension; C: compression; S: shear | |

*5.3 CPU computational performance*

The proposed DJ-TLED was compared against TLED for CPU computational performance using an Intel(R) Core(TM) i5-2500K CPU @ 3.30 GHz with 16 GB RAM PC running Windows 10 operating system without Visual Studio C++ code optimisation. Using the same model in Section 5.1, Fig. 6 presents CPU solution times of a single simulation step in the T4 mesh and H8 mesh using NH, TI, OT, and MR material models for 6 different model sizes. The computation times of DJ-TLED were all less than those of TLED. Using the proposed DJ-TLED, the computation times of NH were the least in both T4 and H8 meshes, followed by those of TI, OT, and MR. It was also observed that the CPU solution times were increased near linearly with the increase of DOFs, and linear interpolation and extrapolation may be used to estimate solution times at unknown model sizes.

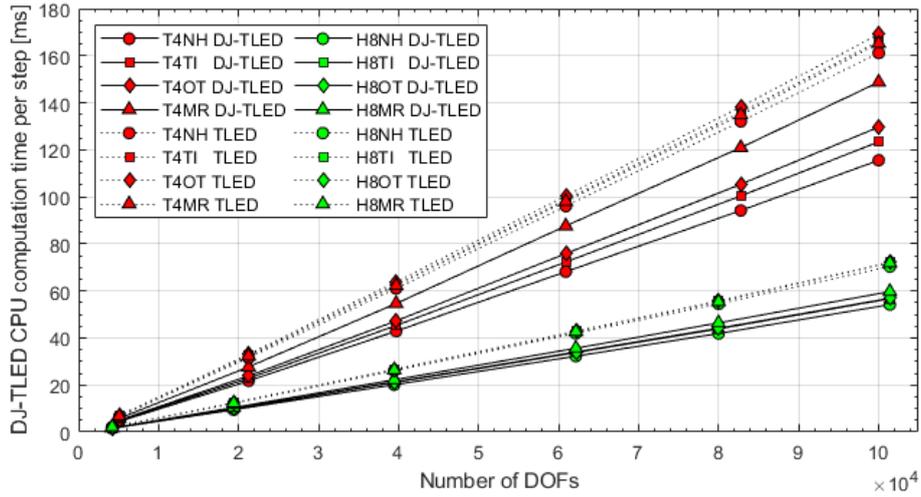

**Fig. 6.** CPU solution times per simulation step.

Fig. 7 presents ratios of computation time of the proposed DJ-TLED over TLED. Table. 5 presents the average speed improvements using DJ-TLED, with the maximum speed improvement of up to 30% achieved with the NH material model in T4 mesh.

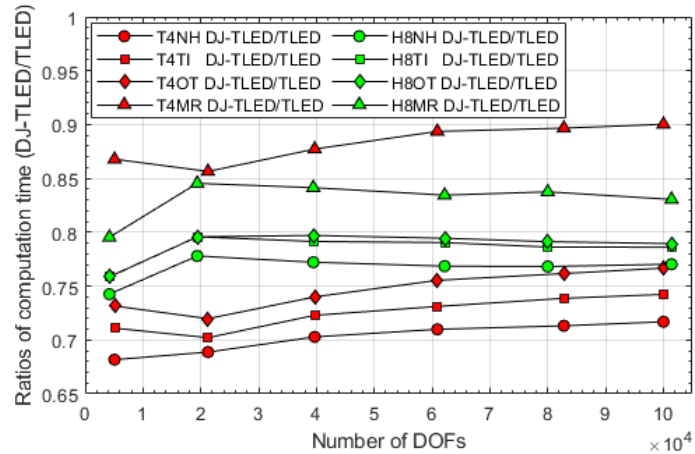

**Fig. 7.** Computation time ratios (DJ-TLED/TLED) in the T4 mesh and H8 mesh using NH, TI, OT, and MR material models at various model sizes.

**Table. 5** Average speed improvements in Fig. 7.

|  |  | T4 | | H8 | |
| --- | --- | --- | --- | --- | --- |
|  |  | Ratios | Improvements | Ratios | Improvements |
| Compared to TLED | NH | 0.70x | 30% | 0.77x | 23% |
|  | TI | 0.73x | 27% | 0.78x | 22% |
|  | OT | 0.75x | 25% | 0.79x | 21% |
|  | MR | 0.88x | 12% | 0.83x | 17% |
| Ratio = DJ-TLED/TLED (x), improvement = (1- DJ-TLED/TLED) * 100% | | | | | |

Fig. 8 further investigates the computational overhead of TI, OT, and MR over NH in the proposed DJ-TLED. Compared to NH, TI incurred additional computation for $\bar{I}_4$ terms, OT incurred additional computation for $\bar{I}_4$ and

$\bar{I}_6$ terms, and MR incurred additional computation for $\bar{I}_2$ terms. The computation of $\bar{I}_2$ terms was more expensive than computing $\bar{I}_4$ or $\bar{I}_6$ terms, which was mainly due to the extra computational operations in $\left( {}^t\hat{\mathbf{g}}^T \cdot \mathbf{I}_{2m} \right)$ (see Eqs. (39-42)). Table. 6 presents the average computational overheads.

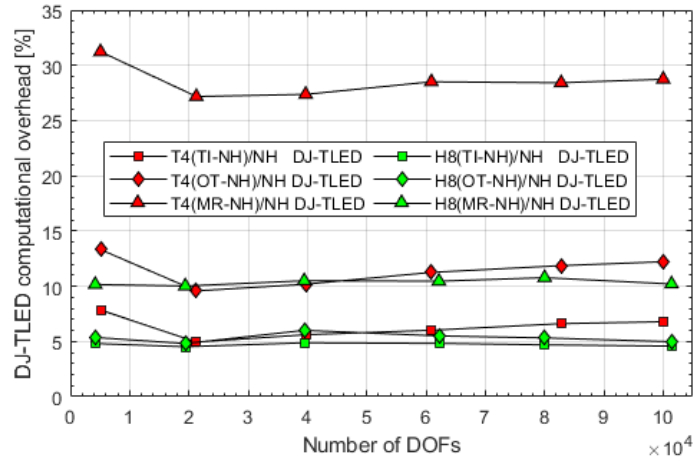

**Fig. 8.** Computational overheads of TI, OT, and MR compared to NH in the proposed DJ-TLED.

**Table. 6** Average computational overheads in Fig. 8.

|  |  | T4 | H8 |
|---|---|---|---|
| Compared to NH | TI | 6.31% | 4.71% |
|  | OT | 11.41% | 5.33% |
|  | MR | 28.58% | 10.35% |

Lastly, Fig. 9 also investigates the element (T4 and H8) computational performance in the proposed DJ-TLED. Linear interpolation was applied to determine the computation times of H8 at T4 model sizes, owing to different DOFs in the T4 and H8 discretisations. For the same DOFs, the proposed DJ-TLED consumed an average between 2.15 and 2.51 times more computation time in T4 than those in H8.

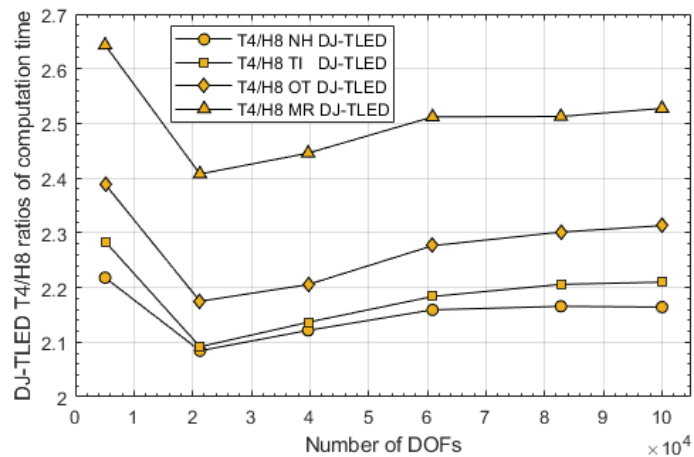

**Fig. 9.** Ratios of computation time (T4/H8) in the proposed DJ-TLED.

## 5.4 GPU computational performance

The GPU implementation was achieved using the NVIDIA CUDA version 10.2 API in C++ and executed using an NVIDIA GeForce GTX 780 GPU on the same PC with the same simulation settings. As presented in Fig. 10, the GPU execution afforded significant speedups over the CPU counterparts, and the solution times also increased almost linearly with the increase of DOFs. Table. 7 presents a summary of GPU speed improvements and computational overheads.

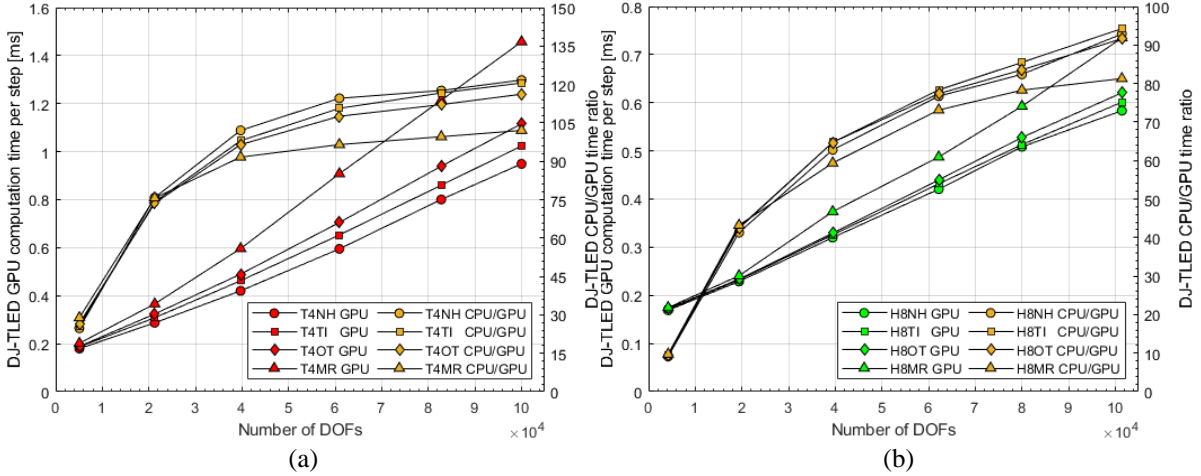

**Fig. 10.** GPU solution times per simulation step and speed improvements in the (a) T4 mesh and (b) H8 mesh using NH, TI, OT, and MR material models.

**Table. 7** GPU speed improvements over CPU and computational overheads in Fig. 10.

|   |   | T4 | | H8 | |
|---|---|---|---|---|---|
| Maximum GPU speed improvements | | | | | |
| Compared to CPU | NH | 121.72x | | 92.85x | |
|  | TI | 120.56x | | 94.26x | |
|  | OT | 116.19x | | 91.63x | |
|  | MR | 102.05x | | 81.29x | |
| GPU computational overheads (range and average) | | | | | |
| Compared to NH | TI | 3.9-9.9% | 7.7% | 0.9-3.0% | 1.9% |
|  | OT | 5.3-18.5% | 14.7% | 2.1-6.4% | 3.7% |
|  | MR | 12.6-53.6% | 39.9% | 3.3-25.9% | 13.9% |

## 5.5 Real-time computational performance

Real-time computation has different performance requirements depending on the application of a simulation. For providing real-time visual feedback of results, the computation speed needs to meet the refresh rate of a monitor (at least $30\ Hz$ or $33.33\ ms$) to achieve continuous motion of the rendered graphics to the human eyes. For providing real-time haptic feedback, the results need to be computed at least $500\ Hz$ ($2\ ms$) to achieve a stable and smooth tactile force rendering from the haptic device [31].

Using extrapolation, Fig. 11 illustrates the model sizes at which the proposed DJ-TLED can provide continuous visual ($30\ Hz$) and haptic ($500\ Hz$) feedback using GPU acceleration on the test computing hardware.

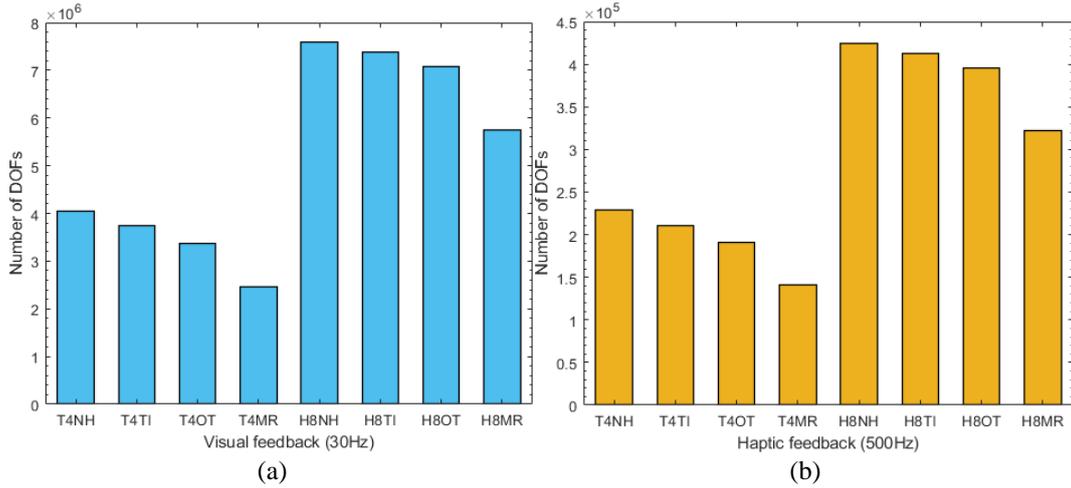

**Fig. 11.** Model sizes for achieving real-time (a) visual and (b) haptic feedback using the proposed DJ-TLED on the test computing hardware.

### 6. Application to neurosurgical simulation of brain deformations

Craniotomy is a surgical procedure to remove a part of the bone from the skull to expose the brain. Due to several physical and physiological processes, it leads to a retraction of the brain surface near the craniotomy (known as brain shift). Due to brain shift, the target positions (such as the positions of brain tumours) are changed from their initial positions in the preoperative images. Acquiring the new positions using intraoperative Magnetic Resonance Imaging (MRI) is often cumbersome, and the scanners are expensive [32]. An alternative approach is to use computer simulation to predict the deformed configuration of the brain to register the preoperative MRIs to the intraoperative brain geometry for safe and reliable neurosurgery. A review of biomechanical modelling of the brain during neurosurgery is presented in Ref. [33].

A simulation of brain deformations for image-guided neurosurgery is presented. The brain model was constructed from a preoperative brain MRI dataset based on RIDER NEURO MRI [34]. The brain surface model was segmented from MRIs using 3D Slicer [35] (http://www.slicer.org) and discretised into a volumetric finite element mesh consisted of 5621 nodes, 16863 DOFs and 28764 T4 elements.

The mechanical behaviour of the brain tissues is considered to be described by an NH hyperelastic constitutive model with mass density $\rho = 1060\ kg/m^3$, shear modulus $\mu = 1006.712\ Pa$, and bulk modulus $\kappa = 50000\ Pa$ (corresponding to Young's modulus $E = 3000\ Pa$ and Poisson's ratio $\nu = 0.49$ [33]). As reported by Wittek *et al.* [36], the predicted brain deformations are only weakly sensitive to the variation of mechanical property values of the brain tissues, and this facilitates the computation of patient-specific biomechanics without the patient-specific material properties of soft tissues, as there are always uncertainties in the identification of tissue properties [32].

A displacement vector $\mathbf{u}(u_x, u_y, u_z)$ = (-9, 9, -7) $mm$ was prescribed to the brain surface near the craniotomy (displacement magnitudes ranged between 10 and 20 $mm$ in the case of a craniotomy-induced brain shift [33]). In the clinical workflow, the prescribed surface displacements can be determined from the rigid registration of the preoperative-to-intraoperative images of the brain. The bottom surface of the brain was assumed to be fixed in position. The brain deformations were computed using DJ-TLED and TLED, where TLED and its alternatives [37] have been used to study the biomechanics of soft tissues. The physical time to be simulated was $t = 10\ s$ with a time step size $\Delta t = 0.00016\ s$ (62500 steps).

Fig. 12 presents the computed brain deformations. As demonstrated in Sections 5.1 and 5.2, the small difference in displacements is due to the small changes in single precision floating-point numbers during different arithmetic operations (see Section 5.1); both methods should intrinsically lead to the same level of numerical accuracy (see Section 5.2). The normalised relative errors (NRE) were calculated by

$$NRE = \left| \frac{u_i^{DJ-TLED} - u_i^{TLED}}{u_{max}^{TLED} - u_{min}^{TLED}} \right| \tag{53}$$

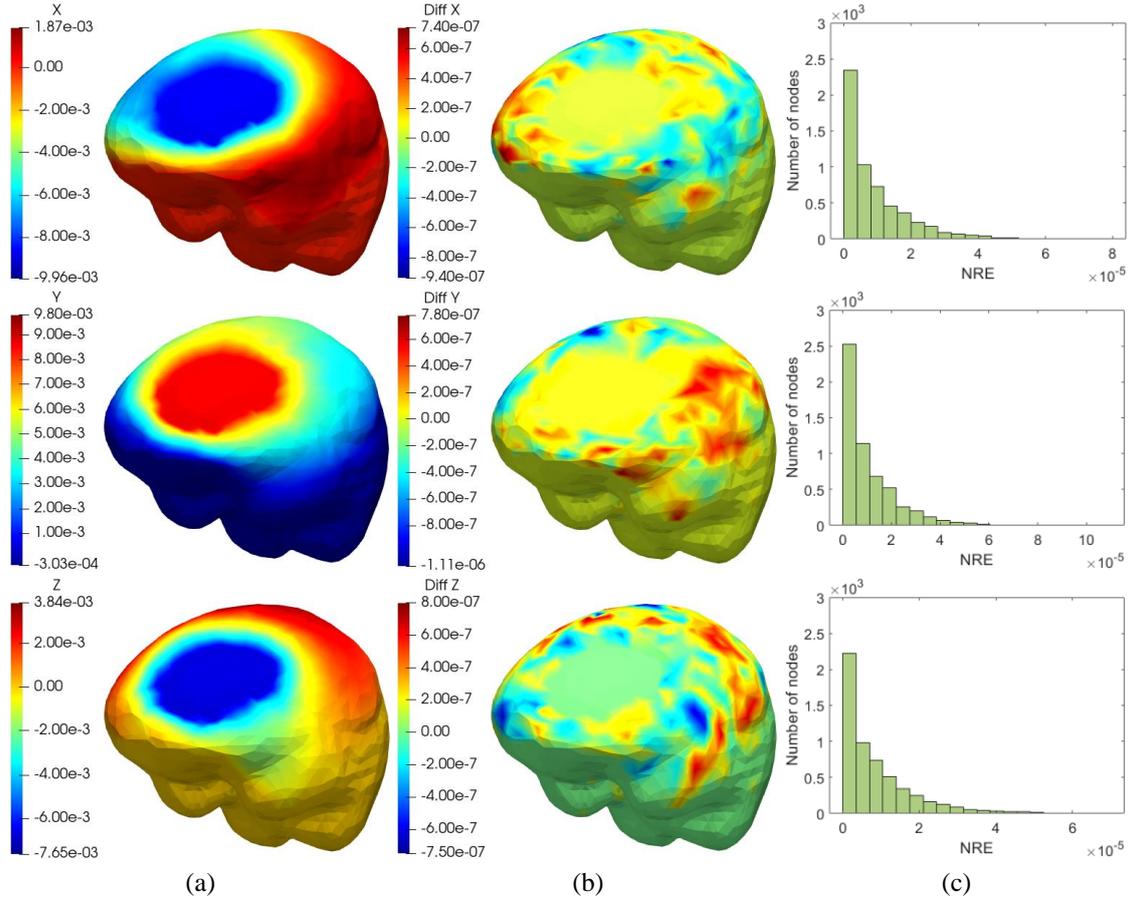

**Fig. 12.** (a) Brain deformations $[m]$ in $x$-, $y$- and $z$-directions computed by the proposed DJ-TLED and (b) compared against TLED reference solutions, (c) histograms show the normalised relative errors, NRE.

Fig. 13 shows the computed displacement field of the brain for a non-rigid transform, which was used to warp the preoperative MRIs to correspond to the brain shift. In intraoperative registration, it requires less than $40\ s$ to provide the surgeon with updated images [38]. In the present simulation, the proposed GPU-accelerated DJ-TLED consumed $15.96\ s$.

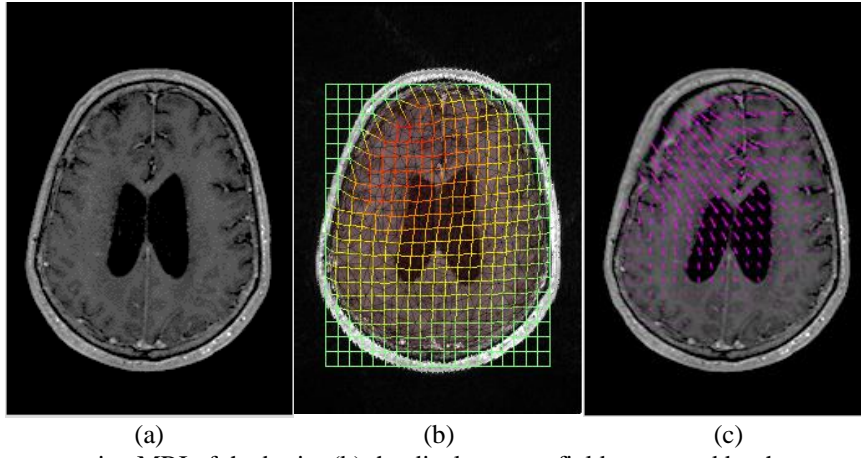

**Fig. 13.** (a) A preoperative MRI of the brain, (b) the displacement field computed by the proposed DJ-TLED is used for image transform, and (c) the transformed preoperative MRI (the arrows show the magnitude and direction of displacements).

In addition to the NH hyperelastic constitutive model (material nonlinearity) and finite-strain brain geometry (geometric nonlinearity), being an explicit solution procedure, the proposed DJ-TLED can also handle geometric nonlinearities such as buckling or wrinkling of the material due to mechanical instabilities. In the computed brain deformations in Fig. 12, some levels of local wrinkling were observed on the brain surface near the craniotomy. In such a mechanical unstable region, the implicit (iterative) scheme may fail to obtain a converged solution or produce a solution with excessive numerical dissipation that in many cases precludes fine, local details of wrinkles. Instead, some works using the explicit solution method, coupled with dynamic relaxation [39,40] for the equilibrium, have been studied to overcome these problems due to geometric instabilities.

## 7. Discussions

The proposed DJ-TLED is compared against the established real-time nonlinear finite element algorithms, TLED and HEML, for similarities and differences. The strengths and limitations are also discussed.

*Comparison with TLED:*

The proposed DJ-TLED is compared against TLED in terms of the force formulation for a simple NH material model. Consider the NH model with strain energy density function $\Psi = \frac{\mu}{2}(\bar{I}_1 - 3) + \frac{\kappa}{2}(J - 1)^2$ given in Table. 3, the corresponding second Piola-Kirchoff stress tensor $^t_0\mathbf{S}$ is given by

$$^t_0\mathbf{S} = J^{-2/3}\mu\mathbf{I} + \left(-\frac{\mu}{3}\bar{I}_1 + \kappa J(J-1)\right)^t_0\mathbf{C}^{-1} \tag{54}$$

The nodal force contributions of the two methods can be obtained by

- *TLED:*

$$_0^t\mathbf{F} = {}_0^t\mathbf{X}\left(J^{-2/3}\mu\mathbf{I} + \left(-\frac{\mu}{3}\bar{I}_1 + \kappa J(J-1)\right){}_0^t\mathbf{C}^{-1}\right){}^0\mathbf{B}\,{}^0V \tag{55}$$

- *DJ-TLED:*

$$_0^t\mathbf{F} = \left(\frac{1}{2}J^{-2/3}\mu\,{}^t\mathbf{J}^T\mathbf{I}_{1m} + \left(-\frac{\mu}{3}\bar{I}_1 + \kappa J(J-1)\right){}^0V\,{}^t\mathbf{J}^{-1}\right)\mathbf{H}_{\xi} \tag{56}$$

Eqs. (55) and (56) share some similarities. The nodal force contributions $_0^t\mathbf{F}$, in both methods, are obtained by multiplying a constant matrix, such as $^0\mathbf{B}$ in TLED and $\mathbf{H}_\xi$ in the proposed DJ-TLED, that consumes the same number of computational operations. In the actual computer implementation, however, the latter can be more computationally efficient in case of the T4 mesh, as the force contribution $_0^t\mathbf{F}_4$ at the fourth node can be computed simply by $_0^t\mathbf{F}_4 = -\sum_{i=1}^{3} {}_0^t\mathbf{F}_i$ in DJ-TLED, which is not the case in TLED.

On the other hand, the main difference occurs in the computation of the time-varying components. In TLED, the deformation gradient $_0^t\mathbf{X}$ is computed by $_0^t\mathbf{X} = \mathbf{I} + {}^t\mathbf{u}\,{}^0\mathbf{B}^T$, and the others are computed by $J = \det(_0^t\mathbf{X})$, $_0^t\mathbf{C} = {}_0^t\mathbf{X}^T\,{}_0^t\mathbf{X}$ and $\bar{I}_1 = J^{-2/3}\text{tr}(_0^t\mathbf{C})$. In the proposed DJ-TLED, by contrast, the Jacobian operator is computed by $^t\mathbf{J} = {}^0\mathbf{J} + {}^t\mathbf{u}\mathbf{H}_\xi^T$, and the others are computed by $J = \det({}^t\mathbf{J})\det({}^0\mathbf{J})^{-1}$ and $\bar{I}_1 = J^{-2/3}\,{}^t\hat{\mathbf{g}}^T\mathbf{m}_1$.

A significant computational difference between DJ-TLED and TLED is that the right Cauchy-Green tensor $_0^t\mathbf{C}$ and its inverse $_0^t\mathbf{C}^{-1}$ are computed at run-time in TLED based on $_0^t\mathbf{C} = {}_0^t\mathbf{X}^T\,{}_0^t\mathbf{X}$, whereas the proposed DJ-TLED computes straightforwardly the Jacobian operator $^t\mathbf{J}$ and its inverse $^t\mathbf{J}^{-1}$. Also, $\mathbf{I}_{1m}$ in DJ-TLED can be precomputed. Table. 8 summarises the difference in computation between DJ-TLED and TLED.

**Table. 8** Differences between the computation in TLED and proposed DJ-TLED.

| Computation at run-time | DJ-TLED | TLED |
|---|---|---|
| $_0^t\mathbf{X}$ | N/A | $\mathbf{I} + {}^t\mathbf{u}\,{}^0\mathbf{B}^T$ |
| $_0^t\mathbf{C}$ | N/A | $_0^t\mathbf{X}^T\,{}_0^t\mathbf{X}$ |
| $_0^t\mathbf{C}^{-1}$ | N/A | $_0^t\mathbf{C}^{-1}$ |
| $^t\mathbf{J}$ | $^0\mathbf{J} + {}^t\mathbf{u}\mathbf{H}_\xi^T$ | N/A |
| $^t\mathbf{J}^{-1}$ | $^t\mathbf{J}^{-1}$ | N/A |
| $\bar{I}_1$ | $J^{-2/3}\,{}^t\hat{\mathbf{g}}^T\mathbf{m}_1$ | $J^{-2/3}\text{tr}(_0^t\mathbf{C})$ |
| $J$ | $\det({}^t\mathbf{J})\det({}^0\mathbf{J})^{-1}$ | $\det(_0^t\mathbf{X})$ |

Note: $^0\mathbf{J}$, $\mathbf{H}_\xi$, $\mathbf{m}_1$ and $\det({}^0\mathbf{J})$ are precomputed in DJ-TLED.

*Comparison with HEML:*

HEML, hyperelastic mass links [14], is also a real-time mesh-based deformation algorithm to simulate hyperelastic deformations. The main similarity between DJ-TLED and HEML is that both algorithms express the right Cauchy-Green tensor $_0^t\mathbf{C}$ by a summation of six individual components. In HEML, $_0^t\mathbf{C}$ is expressed by

$$_0^t\mathbf{C} = l_1^2\mathbf{C}_1 + l_2^2\mathbf{C}_2 + l_3^2\mathbf{C}_3 + l_4^2\mathbf{C}_4 + l_5^2\mathbf{C}_5 + l_6^2\mathbf{C}_6 \tag{57}$$

where $l_1^2, \ldots, l_6^2$ are the six squared edge lengths of a T4 element, and $\mathbf{C}_i = \mathbf{V}^{-T}\mathbf{D}_i\mathbf{V}^{-1}$ (see Ref. [14] for details and notations).

In the proposed DJ-TLED, $_0^t\mathbf{C}$ is expressed by Eq. (19). As such, both algorithms express the strain invariants $I_1$ and $I_2$ in a similar form, i.e.,

- *HEML:*

$$I_1 = \mathbf{vtr} \cdot \mathbf{l} \tag{58a}$$

$$I_2 = \frac{1}{2}\mathbf{l}^T(\mathbf{vtr} \otimes \mathbf{vtr} - \mathbf{Mtr})\mathbf{l} \tag{58b}$$

- *DJ-TLED (Eqs. (27) and (28), they are reproduced here for comparison):*

$$I_1 = {}^t\hat{\mathbf{g}}^T \mathbf{m}_1 \tag{59a}$$

$$I_2 = \frac{1}{2} {}^t\hat{\mathbf{g}}^T[\mathbf{m}_1\mathbf{m}_1^T - \mathbf{W}] {}^t\hat{\mathbf{g}} = {}^t\hat{\mathbf{g}}^T \mathbf{M}_2 {}^t\hat{\mathbf{g}} \tag{59b}$$

However, HEML is limited to modelling only isotropic hyperelastic materials in the T4 mesh. In contrast, the proposed DJ-TLED is established in a general way from TLED that it can be applied to not only T4 but also the H8 mesh and also simulate anisotropic hyperelastic materials. Moreover, real-time computation in the proposed DJ-TLED is achieved using GPU acceleration, which was not demonstrated in HEML.

Lastly, the complete deviatoric terms for a compressible hyperelastic material model were not considered in HEML. Using the strain energy density function $\Psi = C_1(I_1 - 3) - \mu \ln J + \frac{\lambda}{2}(\ln J)^2$ given in HEML [14], it can be written in terms of $\bar{I}_1$ as

$$\Psi = C_1(J^{2/3}\bar{I}_1 - 3) - \mu \ln J + \frac{\lambda}{2}(\ln J)^2 \tag{60}$$

The deviatoric operator $\mathrm{DEV}[\cdot] \equiv [\cdot] - \frac{1}{3}([\cdot]: {}^t_0\mathbf{C}) {}^t_0\bar{\mathbf{C}}^{-1}$ leads to an $\bar{I}_1$ term in the volumetric force component, and it is obtained by

$$-\frac{2}{3}\left(\frac{\partial \Psi}{\partial \bar{I}_1}\bar{I}_1\right) = -\frac{2}{3}C_1 J^{2/3}\bar{I}_1 \tag{61}$$

which was not considered in HEML for a compressible hyperelastic material model.

Due to the above similarities and differences, the proposed DJ-TLED may also be considered as the generalised form of HEML.

*Strengths and limitations:*

The major difference between the proposed DJ-TLED and TLED is that the notions of stress and strain are not explicitly visible in the proposed DJ-TLED but embedded implicitly in the formulation of nodal force contributions. In TLED, the second Piola-Kirchoff stress ${}^t_0\mathbf{S}$ is computed from the right Cauchy-Green tensor ${}^t_0\mathbf{C}$ as a step towards the computation of nodal forces (see Eqs. (4) and (7)). In the proposed DJ-TLED, by contrast, ${}^t_0\mathbf{C}$ does not need to be evaluated explicitly at run-time; instead, only the Jacobian operator ${}^t\mathbf{J}$ is computed. This leads to both the strength and limitation of the proposed method.

Owing to the proposed Jacobian formulation, the computation of the state of stress is not explicitly performed as a solution step in DJ-TLED. As such, if the state of stress is desired, the stresses would need to be computed in a backward fashion from the nodal force contributions (i.e., using nodal internal forces to compute stresses) at the cost of additional computational operations.

On the other hand, it also leads to the strength of the proposed method, in that such a Jacobian formulation can be advantageous for fast deformation computation and is particularly useful if only the displacement field is of interest. For instance, only the final, deformed positions of the brain tissues are interested in the image registration for neurosurgery [41], which is demonstrated in Section 6.

In short, TLED computes the state of stress and then the displacements, whereas DJ-TLED computes the displacements directly and then the state of stress, if needed. This provides DJ-TLED a computational advantage when the displacement field is of primary interest.

Another strength of DJ-TLED is that it can achieve faster computation than TLED while maintaining the same order of numerical accuracy. As demonstrated previously, the proposed DJ-TLED is developed based on TLED by a reformulation of nodal forces to be expressed by using only the Jacobian operator, and hence it achieves faster computation without compromising on numerical accuracy. In contrast, the model order reduction and neural network methods achieve fast computation in a reduced model space where numerical accuracy is dependent on the training data.

As seen from Steps #1-3 in Section 3, the proposed DJ-TLED is constructed based on TLED from the second Piola-Kirchoff stress, which consists of a volumetric stress component and an isochoric stress component. Such a treatment has implications when simulating nearly incompressible hyperelastic materials, such as the brain tissues, where numerical difficulties, often referred to as locking phenomena, are presented due to the over stiffening of the system arise from incompressibility imposed internally. A mitigation of this is to employ the average nodal pressure (ANP) [42], where the volumetric stress component is modified by considering the average of nodal pressures, to reduce the number of incompressibility constraints, and it has been successfully applied to TLED [43]. The ANP technique is not implemented for DJ-TLED in the present work, but it could be constructed from the TLED implementation.

Lastly, it is worth recalling that the proposed DJ-TLED is established in the framework of isoparametric finite element formulation [22], for which the interpolation of the element coordinates and element displacements using the same interpolation functions, defined in the natural coordinate system, is the fundamental basis (Section 2.1.2). As such, some element types using the interpolation functions defined in the local coordinates **x**, rather than the natural coordinates **ξ**, such as the triangular elements constructed using area coordinates in terms of **x**, are not considered. Instead, the area coordinates of the "unit triangle" [22] in terms of **ξ** are used. Similarly, the same treatment is applied to the tetrahedral elements to use the volume coordinates in the natural space.

## 8. Conclusion

In this paper, a direct Jacobian formulation of TLED, named DJ-TLED, is presented, which achieved maximum speed improvements of CPU solution time of up to 30% compared to TLED and up to 121.72× with GPU acceleration using the test computer hardware. The speed improvement is afforded by a reformulation of TLED using only the Jacobian operator, instead of the deformation gradient tensor and finite deformation tensor, for fewer computational operations at run-time. The strength of the proposed DJ-TLED is that it is especially useful when the displacement field is of primary concern, such as in the case of neurosurgical simulation of brain deformations demonstrated. Novel Jacobian formulations are developed for isotropic and anisotropic hyperelastic constitutive material models and evaluated in the 4-node tetrahedral mesh and the 8-node hexahedral mesh with GPU implementation. Quantitative and qualitative comparisons are conducted with state-of-the-art TLED and HEML algorithms for similarities and differences. Overall, this work contributes towards a comprehensive DJ-TLED algorithm for real-time simulation of nonlinear deformation problems. Future research work will investigate the performance of the proposed method when using in conjunction with other methods such as deep learning algorithms for hyperelastic deformations.

## Appendix A. Expression of nodal force contributions ${}_0^t\mathbf{F}$ using Jacobian operator

Recall that the nodal force contributions ${}_0^t\mathbf{F}$ in TLED in Eq. (4) are given by

$$ {}_0^t\mathbf{F} = {}_0^t\mathbf{X}\, {}_0^t\mathbf{S}\, {}^0\mathbf{B}\, {}^0V \tag{A.1} $$

Using the following expressions

$$ {}_0^t\mathbf{X} = {}^t\mathbf{J}^T\, {}^0\mathbf{J}^{-T} \tag{A.2} $$

$$ {}^0\mathbf{B} = {}^0\mathbf{J}^{-1}\mathbf{H}_\xi \tag{A.3} $$

$$ {}_0^t\mathbf{S} = 2J^{-2/3}\frac{\partial \Psi}{\partial \bar{I}_1}\mathbf{I} + \left(-\frac{2}{3}\frac{\partial \Psi}{\partial \bar{I}_1}\bar{I}_1 + J\frac{\partial \Psi}{\partial J}\right){}_0^t\mathbf{C}^{-1} \tag{A.4} $$

Eq. (A.1) can be written as

$$ {}_0^t\mathbf{F} = \left({}^t\mathbf{J}^T\, {}^0\mathbf{J}^{-T}\right)\left(2J^{-2/3}\frac{\partial \Psi}{\partial \bar{I}_1}\mathbf{I} + \left(-\frac{2}{3}\frac{\partial \Psi}{\partial \bar{I}_1}\bar{I}_1 + J\frac{\partial \Psi}{\partial J}\right){}_0^t\mathbf{C}^{-1}\right)\left({}^0\mathbf{J}^{-1}\mathbf{H}_\xi\right){}^0V \tag{A.5} $$

By noting

$$ {}_0^t\mathbf{C}^{-1} = ({}_0^t\mathbf{X}^T\, {}_0^t\mathbf{X})^{-1} = {}^0\mathbf{J}^T\, {}^t\mathbf{J}^{-T}\, {}^t\mathbf{J}^{-1}\, {}^0\mathbf{J} \tag{A.6} $$

Eq. (A.5) can be further written as

$$ {}_0^t\mathbf{F} = \left({}^t\mathbf{J}^T\, {}^0\mathbf{J}^{-T}\right)\left(2J^{-2/3}\frac{\partial \Psi}{\partial \bar{I}_1}\mathbf{I} + \left(-\frac{2}{3}\frac{\partial \Psi}{\partial \bar{I}_1}\bar{I}_1 + J\frac{\partial \Psi}{\partial J}\right){}^0\mathbf{J}^T\, {}^t\mathbf{J}^{-T}\, {}^t\mathbf{J}^{-1}\, {}^0\mathbf{J}\right)\left({}^0\mathbf{J}^{-1}\mathbf{H}_\xi\right){}^0V \tag{A.7} $$

which can be rearranged as

$$ {}_0^t\mathbf{F} = \left(J^{-2/3}\, {}^t\mathbf{J}^T\frac{\partial \Psi}{\partial \bar{I}_1}(2\,{}^0V\,{}^0\mathbf{J}^{-T}\,{}^0\mathbf{J}^{-1}) + \left(-\frac{2}{3}\frac{\partial \Psi}{\partial \bar{I}_1}\bar{I}_1 + J\frac{\partial \Psi}{\partial J}\right){}^0V\,{}^t\mathbf{J}^{-1}\right)\mathbf{H}_\xi \tag{A.8} $$

## Appendix B. Constant matrices $\mathbf{G}_{11}$, $\mathbf{G}_{22}$, $\mathbf{G}_{33}$, $\mathbf{G}_{12}$, $\mathbf{G}_{13}$ and $\mathbf{G}_{23}$

Based on Eq. (16), the constant matrices are defined as

$$\mathbf{G}_{11} = {}^0\mathbf{J}^{-1} \begin{bmatrix} 1 & 0 & 0 \\ 0 & 0 & 0 \\ 0 & 0 & 0 \end{bmatrix} {}^0\mathbf{J}^{-T} \tag{B.1}$$

$$\mathbf{G}_{22} = {}^0\mathbf{J}^{-1} \begin{bmatrix} 0 & 0 & 0 \\ 0 & 1 & 0 \\ 0 & 0 & 0 \end{bmatrix} {}^0\mathbf{J}^{-T} \tag{B.2}$$

$$\mathbf{G}_{33} = {}^0\mathbf{J}^{-1} \begin{bmatrix} 0 & 0 & 0 \\ 0 & 0 & 0 \\ 0 & 0 & 1 \end{bmatrix} {}^0\mathbf{J}^{-T} \tag{B.3}$$

$$\mathbf{G}_{12} = {}^0\mathbf{J}^{-1} \begin{bmatrix} 0 & 1 & 0 \\ 1 & 0 & 0 \\ 0 & 0 & 0 \end{bmatrix} {}^0\mathbf{J}^{-T} \tag{B.4}$$

$$\mathbf{G}_{13} = {}^0\mathbf{J}^{-1} \begin{bmatrix} 0 & 0 & 1 \\ 0 & 0 & 0 \\ 1 & 0 & 0 \end{bmatrix} {}^0\mathbf{J}^{-T} \tag{B.5}$$

$$\mathbf{G}_{23} = {}^0\mathbf{J}^{-1} \begin{bmatrix} 0 & 0 & 0 \\ 0 & 0 & 1 \\ 0 & 1 & 0 \end{bmatrix} {}^0\mathbf{J}^{-T} \tag{B.6}$$

**Appendix C. Constant $\mathbf{m}_1$, $\mathbf{m}_4$, $\mathbf{m}_6$, $\mathbf{M}_2$, $\mathbf{M}_5$ and $\mathbf{M}_7$ defined in the proposed DJ-TLED**

Recall that $\mathbf{m}_1$ in Eq. (23) is defined as

$$\mathbf{m}_1 = [\mathrm{tr}(\mathbf{G}_{11})\ \mathrm{tr}(\mathbf{G}_{22})\ \mathrm{tr}(\mathbf{G}_{33})\ \mathrm{tr}(\mathbf{G}_{12})\ \mathrm{tr}(\mathbf{G}_{13})\ \mathrm{tr}(\mathbf{G}_{23})]^T \tag{C.1}$$

Similarly, $\mathbf{m}_4$ and $\mathbf{m}_6$ concerning the preferred fibre direction matrices $\mathbf{A}$ and $\mathbf{B}$ can be expressed as

$$\mathbf{m}_4 = [\mathrm{tr}(\mathbf{A}\mathbf{G}_{11})\ \mathrm{tr}(\mathbf{A}\mathbf{G}_{22})\ \mathrm{tr}(\mathbf{A}\mathbf{G}_{33})\ \mathrm{tr}(\mathbf{A}\mathbf{G}_{12})\ \mathrm{tr}(\mathbf{A}\mathbf{G}_{13})\ \mathrm{tr}(\mathbf{A}\mathbf{G}_{23})]^T \tag{C.2}$$

$$\mathbf{m}_6 = [\mathrm{tr}(\mathbf{B}\mathbf{G}_{11})\ \mathrm{tr}(\mathbf{B}\mathbf{G}_{22})\ \mathrm{tr}(\mathbf{B}\mathbf{G}_{33})\ \mathrm{tr}(\mathbf{B}\mathbf{G}_{12})\ \mathrm{tr}(\mathbf{B}\mathbf{G}_{13})\ \mathrm{tr}(\mathbf{B}\mathbf{G}_{23})]^T \tag{C.3}$$

Recall that $\mathbf{W}$ in Eq. (25) is defined as

$$\mathbf{W} = \begin{bmatrix} \mathrm{tr}(\mathbf{G}_{11}{}^T\mathbf{G}_{11}) & \mathrm{tr}(\mathbf{G}_{11}{}^T\mathbf{G}_{22}) & \cdots & \mathrm{tr}(\mathbf{G}_{11}{}^T\mathbf{G}_{23}) \\ \mathrm{tr}(\mathbf{G}_{22}{}^T\mathbf{G}_{11}) & \mathrm{tr}(\mathbf{G}_{22}{}^T\mathbf{G}_{22}) & \cdots & \vdots \\ \vdots & \vdots & \ddots & \vdots \\ \mathrm{tr}(\mathbf{G}_{23}{}^T\mathbf{G}_{11}) & \cdots & \cdots & \mathrm{tr}(\mathbf{G}_{23}{}^T\mathbf{G}_{23}) \end{bmatrix} \tag{C.4}$$

which is used to define $\mathbf{M}_2$ as

$$\mathbf{M}_2 = \frac{1}{2}[\mathbf{m}_1\mathbf{m}_1{}^T - \mathbf{W}] \tag{C.5}$$

Similarly, $\mathbf{M}_5$ and $\mathbf{M}_7$ concerning the preferred fibre direction matrices $\mathbf{A}$ and $\mathbf{B}$ can be expressed as

$$\mathbf{M}_5 = \begin{bmatrix} \mathrm{tr}(\mathbf{A}\mathbf{G}_{11}{}^T\mathbf{G}_{11}) & \mathrm{tr}(\mathbf{A}\mathbf{G}_{11}{}^T\mathbf{G}_{22}) & \cdots & \mathrm{tr}(\mathbf{A}\mathbf{G}_{11}{}^T\mathbf{G}_{23}) \\ \mathrm{tr}(\mathbf{A}\mathbf{G}_{22}{}^T\mathbf{G}_{11}) & \mathrm{tr}(\mathbf{A}\mathbf{G}_{22}{}^T\mathbf{G}_{22}) & \cdots & \vdots \\ \vdots & \vdots & \ddots & \vdots \\ \mathrm{tr}(\mathbf{A}\mathbf{G}_{23}{}^T\mathbf{G}_{11}) & \cdots & \cdots & \mathrm{tr}(\mathbf{A}\mathbf{G}_{23}{}^T\mathbf{G}_{23}) \end{bmatrix} \tag{C.6}$$

$$\mathbf{M}_7 = \begin{bmatrix} \text{tr}(\mathbf{BG}_{11}{}^T\mathbf{G}_{11}) & \text{tr}(\mathbf{BG}_{11}{}^T\mathbf{G}_{22}) & \cdots & \text{tr}(\mathbf{BG}_{11}{}^T\mathbf{G}_{23}) \\ \text{tr}(\mathbf{BG}_{22}{}^T\mathbf{G}_{11}) & \text{tr}(\mathbf{BG}_{22}{}^T\mathbf{G}_{22}) & \cdots & \vdots \\ \vdots & \vdots & \ddots & \vdots \\ \text{tr}(\mathbf{BG}_{23}{}^T\mathbf{G}_{11}) & \cdots & \cdots & \text{tr}(\mathbf{BG}_{23}{}^T\mathbf{G}_{23}) \end{bmatrix} \quad (C.7)$$